\documentclass{aa}
\usepackage{graphicx} 
\usepackage{txfonts}
\usepackage[dvips]{epsfig}
\usepackage{natbib}
\bibpunct{(}{)}{;}{a}{}{,} 

\newcommand{\um}{\,$\mu$m}
\newcommand{\uJy}{\,$\mu$Jy}
\newcommand{\LIR}{L$_\mathrm{IR}$}
\newcommand{\LIRmath}{L_\mathrm{IR}}
\newcommand{\Lsun}{L$_\odot$}
\newcommand{\LFIR}{LF$_\mathrm{IR}$}

\newcommand{\M}[1]{{\mathbf{#1}}}		
\newcommand{\Eq}[1]{{Eq.~(\ref{e:#1})}}	
\newcommand{\Eqs}[1]{{Eqs.~(\ref{e:#1})}}	
\newcommand{\Ep}[1]{{~(\ref{e:#1})}}	
\def\EQN#1{\label{e:#1}}  
\def\T#1{{{{#1}}^{\bot}}}		
\def\d#1{{{\rm d}{#1}}}		
\def\mdot{\!\cdot\!}		
\def\R#1{{\mathrm{#1}}}         

\begin{document}

\title{Cosmic star-formation history \\ from a non-parametric inversion of infrared galaxy counts}

\authorrunning{Le Borgne et al.} 
\titlerunning{Cosmic star-formation history from galaxy counts}

\author{D. Le Borgne\inst{1,2}, D. Elbaz\inst{1}, P. Ocvirk\inst{1,3},
  C. Pichon\inst{2}}

\offprints{Damien Le Borgne, \email{leborgne@iap.fr}}

\institute{CEA/Saclay, DSM/IRFU/SAp, F-91191 Gif-sur-Yvette, France
\and  Institut d'Astrophysique de Paris, UMR7095 CNRS, UPMC, 
98bis boulevard Arago, F-75014 Paris, France
\and Astrophysikalisches Institut Potsdam, An der Sternwarte 16,
D-14482 Potsdam, Germany
}

\date{Received...; accepted... }

\abstract  
{}
{This paper aims at providing new
conservative constraints on the cosmic star-formation (SF) history from the
empirical modeling of recent observations in the mid and far infrared. }
{We present a new empirical method based on a non-parametric inversion
technique. It primarily uses multi-wavelength galaxy counts in the
infrared and sub-mm (15, 24, 70, 160, 850 $\mu$m), and it does not require
any redshift information. 
This inversion can be considered as a ``blind'' search for all possible evolutions and shapes
of the infrared luminosity function of galaxies, from which the
evolution of the star-formation rate density (SFRD) and its uncertainties are derived. The cosmic
infrared background (CIRB) measurements are used a posteriori to
tighten the range of solutions. The inversion relies only on two
hypotheses: (1) the luminosity function remains smooth both in
redshift and luminosity, (2) a set of infrared spectral energy
distributions (SEDs) of galaxies must be assumed, with a dependency on the total
luminosity alone.}
{The range of SF histories recovered at low redshift is
well-constrained and consistent with direct measurements from various
redshift surveys. Redshift distributions are recovered without any
input into the redshifts of the sources making the counts. A peak of the
SFRD at $z\simeq 2$ is preferred, although higher redshifts are not
excluded. We also demonstrate that galaxy counts at 160\,$\mu$m
present an excess around 20 mJy that is not consistent with counts at
other wavelengths under the hypotheses cited above. Finally, we find 
good consistency between the observed evolution of the stellar mass
density and the prediction from our model of SF history.}
{Multi-wavelength counts and CIRB (both projected observations) alone,
interpreted with a luminosity-dependent library of SEDs, contain
enough information to recover the cosmic evolution of the infrared
luminosity function of galaxies, and therefore the evolution of the SFRD, with quantifiable
errors.  Moreover, the inability of the inversion to model perfectly
and simultaneously the multi-wavelength infrared counts implies either
(i) the existence of a sub-population of colder galaxies, (ii) a
larger dispersion of dust temperatures among local galaxies than
expected, (iii) a redshift evolution of the infrared SED of galaxies.}


\maketitle

\section{Introduction}

Some key questions remain concerning the formation of galaxies, such
as when and how galaxies formed their stars over the past
13~Gyr. Thanks to recent ultra-deep surveys at various wavelengths,
some phenomena are now quite accurately measured and described, at
least at relatively low redshift.  For instance, it is well-established that massive galaxies have experienced most of their SF activity at early epochs, whereas the SF activity in
small galaxies keeps a more constant level, on average. This so-called
``downsizing'' has been subject to many studies over the past few
years
\citep{MadFerDic96,LilLeFHam96,SteAdeGia99,JunGlaCra05,LeFPapDol05}
and various signs of this downsizing are now seen. But precise
measurements of the rate of stellar formation occurring at high
redshift are still needed to efficiently challenge the latest models
of galaxy formation. In other words, additional constraints on the
modeling of the evolution of the cosmic SF history should
be inspired by observations.

Recently, very deep surveys were designed to probe SF in the distant
universe.  For instance, mid-infrared light (at 15 and 24\um{}) collected by the
ISO and Spitzer telescopes has been used extensively to measure the
star-formation rate (SFR) of both nearby and distant galaxies. The
evolution of the infrared (IR) luminosity functions (hereafter LF),
parameterized in shape (e.g. with Schechter functions), has
been measured up to $z=2$ with, again, a parameterization for the
evolution that can be both in luminosity ($(1+z)^{\alpha_L}$) or in
density ($(1+z)^{\alpha_D}$). From these studies
\citep{LeFPapDol05,BabRowVac06,CapLagYan07}, several values of $\alpha_L$ and
$\alpha_D$ have been measured in various redshift ranges and used to
derive the evolution of the SFRD. Such works give a 
very solid basis to our understanding of galaxy evolution, but they
are generally limited to relatively low redshifts for two
reasons. First, they are very expensive in observation time if
spectroscopic redshifts are used to derive the luminosities of the
sources. Photometric redshifts can also be used to complement
spectroscopic redshifts, but their uncertainties are well-quantified
only at low redshift, where spectroscopic redshifts are available to
calibrate them. The second reason is that k-corrections of 24\um{}
light becomes large  and hazardous at $z>2$ where the
restframe wavelength falls in the PAH features.  The derivations of
total infrared luminosities and SFRs are
therefore uncertain.  In the following, we call this approach ``direct
method'' and it consists in deriving luminosity functions
and SFRD from mid-IR light collected in redshift surveys.

However, these studies present some severe limitations.
First, a full multi-wavelength approach has not yet been used to \emph{measure}
these quantities. Indeed, the studies cited above only use observations
in one mid-IR band to extrapolate to a total IR luminosity and derive an
SFR from uncertain calibrations \citep[e.g.][]{Ken98}.
Moreover, various depths and areas must be explored simultaneously.
On the one hand, only very deep surveys are able to probe
the lowest levels of SF in distant galaxies, which is necessary
to account for the total volume-average SF activity potentially
dominated by numerous galaxies with low SFRs. But these very deep surveys
necessarily probe only a small area in the sky. On the other hand, large (and therefore
shallow) surveys are also needed to probe the populations of
sources presenting a low density on the sky. It is the case for very
low-redshift sources and it might also be the case for the distant
populations of
ULIRGS\footnote{We adopt the following notation in this paper:\\
``Normal'' galaxies have $\LIRmath{}<10^{11} L_\odot$ \\ ``LIRGS'':
$10^{11} L_\odot < \LIRmath{} \leq 10^{12} L_\odot$ \\ ``ULIRGS'':
$10^{12} L_\odot < \LIRmath{} \leq 10^{13} L_\odot$ \\ ``HLIRGS'':
$\LIRmath{}>10^{13} L_\odot$ \\} for instance.

A natural way of exploiting this multi-wavelength and
multi-scale information all-together is to adopt a global modeling
approach. Ideally, the models (defined as the combination of a library
of IR SEDs and of an evolving LF) must be able to simultaneously account
 for all the counts observed at all IR wavelengths, from
faint to bright sources. In addition, they must also account for the
constraints brought by measurements of the CIRB, and from LFs measured
with the ``direct method''.  For instance, \citet{ChaElb01} (hereafter CE01) and
\citet{LagDolPug03} (hereafter LDP03) or \citet{FraAusCes01} have found
models that
are able to reproduce most of these
constraints. However, some adjustments of the LFs or even of the SEDs
by these authors were needed to reproduce the most up-to-date
observations. This modeling approach is powerful but it is also
subject to caveats. Indeed, some important choices must be made for how
to parameterize the shape (e.g. a double power law, a Schechter function, or the local
15\um{} LF from \citet{Xu00} converted into \LIR{}) and the evolution of
the LFs (e.g. with factors $(1+z)^{\alpha_L}$ and
$(1+z)^{\alpha_D}$). These parameterizations rely mainly on physical
intuition and sometimes require adding more degrees of freedom. This
is the case for the two populations of sources introduced by LDP03,
with their LFs evolving separately from each other. Moreover, these
models cannot claim to be the only possible representation of the true
cosmic LFs or of the true SEDs. They are generally good enough to reproduce
current observations, but they are never provided with a range of
uncertainties.

In this paper, we add new constraints to the decomposition of the
``Lilly-Madau'' SFRD diagram using a powerful non-parametric
inversion technique that blindy and \emph{simultaneously} exploits  the information
from the published multi-wavelength IR galaxy counts in deep and small, as
well as large and shallow, surveys. From these counts alone, and without
any input information on redshifts, we derive the {\it range} of
\emph{all} possible evolutions and shapes of the IR LF.


In Sect.~\ref{section:inversion}, we present the inversion method in
the general case (see also Appendix~\ref{s:NP}). We present in
Section~\ref{section:libraries} the data and our choice for an SED
library used in this study. Section~\ref{section:results} contains our
results: the counts inversion and the corresponding IR LFs, together
with the inferred cosmic SF history. In
Section~\ref{section:validation}, we validate our inversion by comparing
our empirical modeling of counts, LFs, SFRD, and stellar-mass density
evolution to bibliographic data (see also
Appendix~\ref{section:robustness} for the robustness of the inversion).
Finally, we discuss the results and give our conclusions in
Section~\ref{section:conclusion}.  Some predictions for forthcoming
Herschel observations are also given in
Appendix~\ref{section:predictions}.  We use a
cosmology defined by H$_{0}$= 70 km s$^{-1}$ Mpc$^{-1}$, $\Lambda$= 0.7,
$\Omega_{\rm m}$=0.3. The IMF is assumed to be \citet{Sal55} unless
otherwise stated (i.e. in Sect.~\ref{section:mstar}).

\section{Method: non-parametric inversion of deep multi-$\lambda$ IR galaxy counts}
\label{section:inversion}

In this section, we present a new approach to infer the evolution of IR
LFs of galaxies from multi-wavelength and multi-scale observations of
galaxy counts.
It consists in a phenomenological modeling approach, similar to the
works of CE01, LDP03, and others\footnote{This approach is at variance
  with the classical ``direct'' methods used for this purpose, in
  which luminosities of individual sources are derived from their
  measured redshifts and fluxes in a {\it single} mid-IR band.}.
However, the modeling is made here in a global and flexible way:
assuming that a given library of SEDs is able to account for the
spectra of galaxies at any redshift (an assumption tested
from the limits of the method's success), we search blindly for all
shapes and evolutions of the total IR LF that are able to reproduce
the multi-wavelength IR counts and the CIRB. This method is
\emph{non-parametric}; i.e., it does not depend on a parameterization
of the LF (in shape or evolution). It must be noted, however,
  that the underlying model of SEDs (on which the inversion depends) can,
  itself, involve one or several parameters that must be fixed for the inversion.

Our method exploits data from infrared surveys designed to probe
high-redshift populations by using their observed galaxy number
counts. However, the redshifts of the sources are not required, which
makes this method quite versatile\footnote{It is worth noting that we can take
(optionally and when available) the redshift information into account
by using as priors the luminosity functions measured at low redshifts
from direct methods. We will show in Sect.~\ref{section:results}
that this knowledge of the luminosity-redshift distribution at low
redshift ($z<2$) brings actually little new to our results.}.

An important advantage of this technique, one that makes it different
from all previous models for IR galaxy number counts, is that it provides an automatic, hence objective,
way of sampling the range of possible histories of the IR luminosity
function.  While previous studies have always presented their favorite
model for fitting the IR galaxy counts, the present work spans the range of
all possible evolutions that are consistent with the observations.
This results in two major improvements. First, the modeled cosmic SFR
history or luminosity functions per redshift bin are presented with
their error bars. Second, it will allow us to discuss the limitation
of local IR SEDs at reproducing the properties of distant
galaxies. Indeed, if after spanning all possibilities, we still find
that the fit is not complete, this will demonstrate that the IR SEDs
must be revised, either because they provide an incomplete description
of local galaxies or because they evolve with redshift.

In this section, as well as in appendix~A, we present the technical and mathematical
aspects of our counts modeling.

\subsection{Linear matrix modeling of the counts}
The deep galaxy counts can be seen as the projection on a flux scale
of the SEDs of galaxies of various luminosities, masses, types,
etc., distributed in redshift. Therefore, to reproduce counts, one
needs at least a description of the number of galaxies at
various redshifts per unit volume, the
SEDs of these galaxies\footnote{In this subsection presenting the
formalism in the general case, an evolution of the SEDs is considered as
possible.}, and a cosmology.

In the following, to describe the numbers of galaxies in space and
time, we use total infrared luminosity functions (\LFIR{}). Doing so,
we simultaneously assume that the SED of a galaxy seen in the IR can
be efficiently described by the sheer knowledge of its redshift and
\LIR{}. Although such a description can be regarded as simple-minded,
the current knowledge of galaxy SEDs in the IR is not much better than
an empirical description parameterized only by \LIR{}. Therefore, the model
description that we have chosen is suited to the current limitations of our
understanding.

For a galaxy (which IR SED is known) lying at redshift $z_0$ with a
luminosity $L^\mathrm{IR}_0$, we can easily compute a k-correction and a
distance modulus to obtain the flux density $S_0$ that one would measure
in a given IR filter centered at the wavelength $\lambda_0$. This
conversion only depends on the cosmology and on the SED of the galaxy.

A luminosity function being only the description of the number of
sources per comoving Mpc$^3$ as a function of redshift and infrared
luminosity \LIR{}, we can then compute a matrix that converts this
evolving \LFIR{} into numbers of galaxies seen at various fluxes and
through different IR filters. This matrix is simply the response
function of the conversion from $(z, \LIRmath)$ to $(S,\lambda)$
presented above. Again, this matrix depends only on the cosmology and
on the library of SEDs. Since it includes k-correction, 
distance effects (dimming as a function of the square of the luminosity
distance), and redshift effects (flux density stretching and dimming), 
we call it the ``k+d'' matrix in the
following\footnote{If the SEDs vary not only with \LIR{} but
  also with $z$, we could call it the ``e+k+d'' matrix.}.

After discretization, and using a matrix notation, the inverse problem can be formalized as

\begin{equation}
  \mathbf{Y}(\lambda,S) = \mathbf{M}(\lambda,S,z,L_\mathrm{IR}) \cdot \mathbf{X}(z,L_\mathrm{IR})\,,
  \label{eq:YMX}
\end{equation}
where $\mathbf Y$ is the matrix containing the number counts at fluxes
$S$ in bands $\lambda$, $\mathbf M$ is  the above-mentioned
``k+d''  transformation matrix, and $\mathbf X$ is
the \LFIR{} that is a function of only $z$ and \LIR{}  (see Appendix~A for details).

Therefore, our problem involves inverting this linear equation to find
the evolving LFs (i.e. redshift dependent number counts per unit volume), $\mathbf X$,
from the known values of $\mathbf Y$ (the wavelength dependent
observed number counts per unit area). Because the matrix, $\mathbf M$, is not
square, and because the number counts are noisy and must be
positive, the solution is not quite as simple as using the pseudo
inverse: $\mathbf{X} = \mathbf{M}^{(-1)} \cdot \mathbf{Y}$, and
requires computing a regularized solution as discussed bellow and
explained in Appendix~A. The reader may also refer to, {\sl e.g.}
\citet{PicSieBie02} or \citet{OcvPicLan06}.  The
uncertainties on the observed counts $\mathbf Y$ are taken into
account through an additional error matrix $\mathbf W$ that makes it
possible to compute a $\chi^2$ between the model $\mathbf X$ and the
data $\mathbf Y$ and to derive uncertainties on the recovered LF.

In this formalism, we choose to describe the luminosity
function in a logarithmic scale, taking $\log_{10}$\LIR{}
instead of \LIR{} everywhere. This is justified by the better conditioning of
the inversion in this case because the \LFIR{}  generally spans several
orders of magnitudes in luminosity.  For the same reason, the counts
are treated numerically through their Euclidian differential form $\d
N/\d S \times S^{2.5}$ (units of mJy$^{1.5}$~deg$^{-2}$) which varies
slowly with flux.  The evolution is measured as a function of
$\log_{10}(1+z)$. Indeed, in the following, we impose a smooth
evolution in redshift of the luminosity function and we need to define
the temporal parameter on which this smoothing applies. We find that
the simplest redshift description that corresponds roughly to a
regular time sampling is actually $\log_{10}(1+z)$. A parameterization
with $z$ or with $\log_{10} z$ would leave too much room for 
strong variations in \LFIR{} at early and late times, respectively.
More quantitatively, we discretize the problem in bins regular in $\delta \log_{10}$\LIR{}=0.1
and in $\delta \log_{10} (1+z)=0.015$, which corresponds to the sampling that
is good enough to produce counts with  regular flux sampling of
$\delta \log_{10}$ S{}=0.1.

\subsection{Regularizing the inverse problem}

In practice, provided the number of bins in \LFIR{} is large enough,
 this problem is ``ill-posed'': there is possibly a large
 number of LFs that are able to satisfy Eq.~(\ref{eq:YMX}) perfectly,
 hence overfitting the noisy counts. However, many of these solutions
 are unlikely and not physical so that we need to ``regularize'' the
 problem to obtain valid solutions. The first natural constraint is
 the positivity of the \LFIR{} (there are no such things as negative
 numbers of galaxies). This imposes an iterative, CPU-costly approach
 to the problem, and a choice for an initial guess (see below).
 Moreover, we want to avoid solutions that are not meaningful given
 our noisy finite set of counts, such as LFs that are chaotically
 varying as a function of \LIR{} or redshift. Therefore, we penalize
 the inversion with an extra term, added to the formal $\chi^2$, which
 enforces the smoothness of the \LFIR{} both in \LIR{} and in $z$
 \citep[see Appendix~A and ][for the details of the
 formalism]{OcvPicLan06}. With these constraints, the
 solutions, $\mathbf X$ (which depend on the choice of the initial
 guess required for the nonlinear optimization), are reasonable and
 can serve as a solid basis for future works. We can also optionally impose external
 constraints as supplementary priors, namely the low-redshift luminosity
 functions obtained from direct methods.

To obtain the range of all realistic solutions for the nonlinear
 optimization problem, we explore a wide range of random initial
 guesses, in a Monte-Carlo approach with at least 100 realizations. The
 range of LFs spanned by the initial guesses is much wider than the
 range of the final \LFIR{} ($X$) obtained after convergence, which
 lends credibility to our study's completeness\footnote{We could also
 have used a more classical approach to characterize the uncertainties
 in the recovered LFs by computing the posterior variance covariance of
 the parameters away from the itterative nonlinear solution, and
 compared it to the corresponding initial prior (hence constructing the
 so called information matrix). These uncertainties naturally arise from
 the uncertainties on the observed counts through the error matrix
 $\mathbf W$. Instead, we favor here Monte-Carlo simulations because
 they yield similar -- if not more robust (since we span ranges of
 possible nonlinear solutions) -- constraints on the recovered LF.}.

Finally, some of the solutions do not match the constraints brought by
CIRB measurements. We filter out these invalid solutions, a posteriori,
leaving only the most realistic luminosity functions.

\section{Data and SEDs used in this work}
\label{section:libraries}
\subsection{Input multi-wavelength counts}

  \begin{figure*}
    \includegraphics[width=0.99\textwidth]{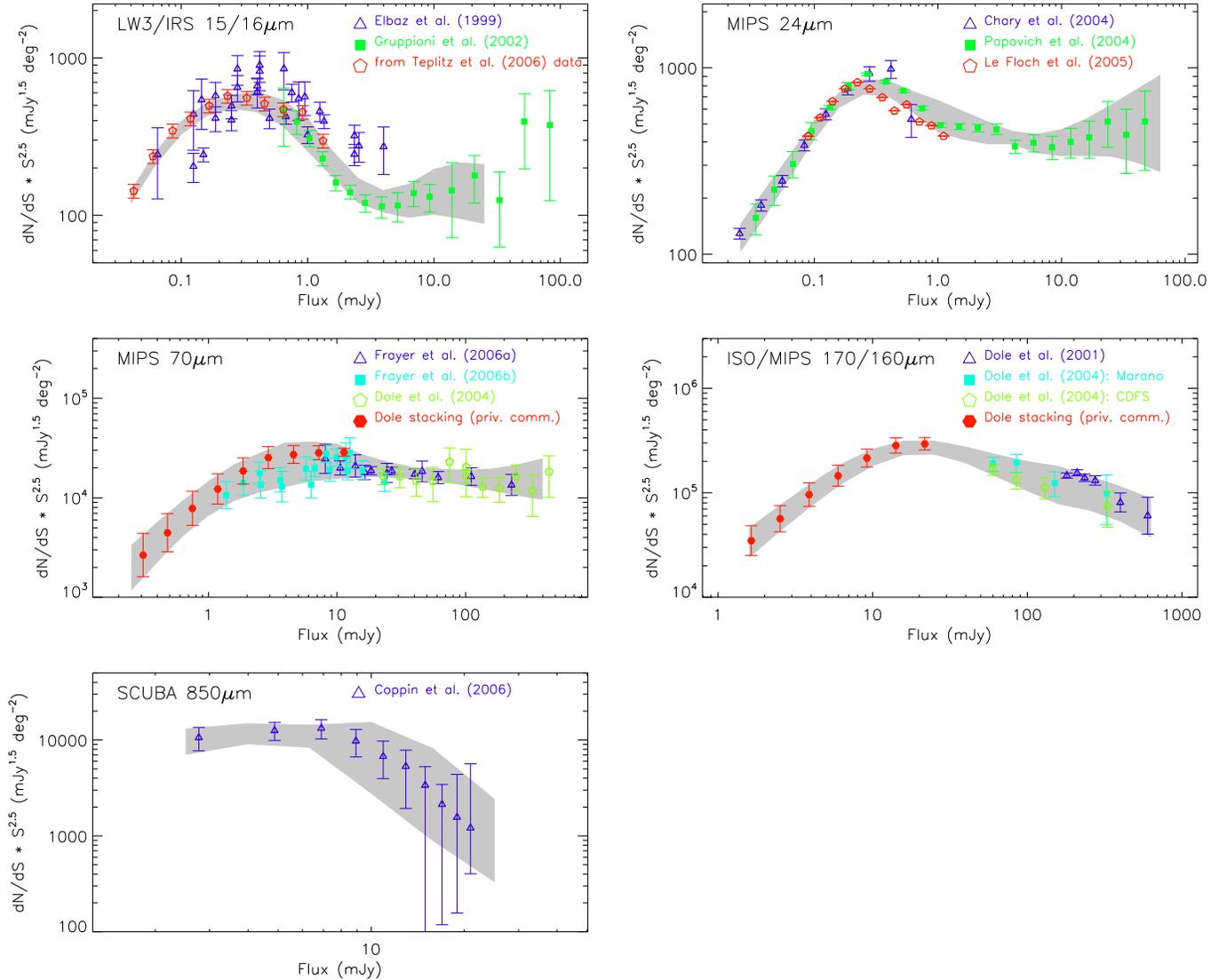}
    \caption{Observed 15/16\um{}, 24\um{}, 70\um{}, 160\um{}, and
    850\um{} counts of galaxies from various surveys. The shaded area
    representing $1~\sigma$ uncertainty area are determined
    semi-automatically from the data points (smoothed envelopes of the data points after excluding
    the strongest outliers, particularly at 15\um{}), and are used as
    an input for the inversions. Instruments/fields: ISO LW3/various
    fields \citep{ElbCesFad99}, {\sc ISO LW3/ELAIS-S}
    \citep{GruLarPoz02}, {\sc Spitzer IRS/GOODS} \citep{TepChaCol06},
    {\sc Spitzer MIPS24/GOODS} \citep{ChaCasDic04}, {\sc Spitzer
    MIPS24/}various fields \citep{PapDolEga04}, {\sc Spitzer
    MIPS24/GOODS} \citep{LeFPapDol05}, {\sc Spitzer MIPS70/FLS}
    \citep{FraFadYan06}, {\sc Spitzer MIPS70/GOODS-N}
    \citep{FraHuyCha06}, {\sc ISO ISOPHOT170/FIRBACK}
    \citep{DolGisLag01}, {\sc Spitzer MIPS160/Marano+CDFS}
    \citep{DolLeFPer04}, {\sc JCMT SCUBA/SXDF+LH} \citep{CopChaMor06}.
   }
    \label{figure:obsdata}
  \end{figure*}

  The multi-wavelength counts that we invert in this work are presented
  in Fig.~\ref{figure:obsdata}. They summarize bibliographic observed
  galaxy counts at 15, 24, 70, 160, and 850\um{}. This compilation is
  not exhaustive but we retained significant surveys for each
  IR band, mixing wide and shallow ones to avoid cosmic variance effects
  as much as possible, and deep ones for faint source detections. The
  data shown in Fig.~\ref{figure:obsdata} are actual observations,
  except for the 70 and 160\um{} points from H. Dole \citep[private
  communication, and also][]{DolLagPug06}, which are obtained from
  stacking 24\um{} sources in several flux bins and using an MIR-FIR
  observed correlation. Although these points are not strict
  measurements, the MIR-FIR relationship is tight enough (down to less
  than 0.1~mJy) to be confident in the stacking results.  We group the
  15 and 16\um{} counts, as well as the 160/170\um{} counts, thus
  neglecting the small differences to the counts (a few percent) that
  are caused by the slightly different k-corrections. Our inversion is
  applied to this compilation.

\subsection{Library of SEDs}

The IR SEDs that we use in this study are taken from the empirical library of CE01,
which defines a bijection between \LIR{} and the IR SED from 3 to 1000
\um{}.  Although this library is based on correlation observed in local
galaxies, we suppose in the following that the SEDs of galaxies with a
given \LIR{} do not change with redshift. It does not necessarily mean
that SEDs of individual galaxies do not evolve, but that they must evolve along the local SED-\LIR{}
correlation. It is worth noting that when a description of the evolving IR SEDs of
galaxies becomes available ({\sl e.g.} with Herschel), the technique
that we use can be applied to these evolving SED libraries.

\begin{figure}
    \includegraphics[width=0.49\textwidth]{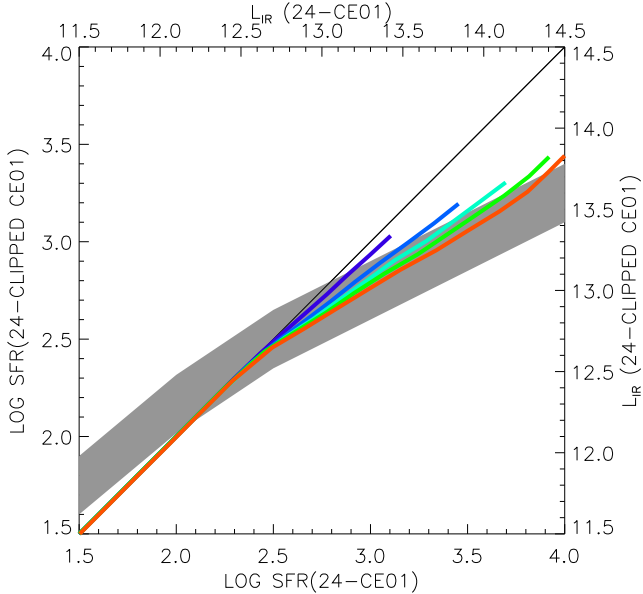}
    \caption{24\um{} vs total IR luminosities (or the SFR equivalent
      of these luminosities assuming a Salpeter IMF). The shaded gray
      area represents the observations of \citet{PapRudLeF07} Fig. 11,
      considering that most of this intense SF is probed by the IR
      light, neglecting the UV contribution. The solid lines
      correspond to the ``clipped'' CE01 library for a set of virtual
      galaxies at redshifts z=1.5, 1.8, 2.2 and 2.5 (from left to
      right) with fluxes fainter than 1~mJy, for which the
      \citet{PapRudLeF07} observations are valid. A pure CE01 library
      would have followed the one-to-one slope.}
    \label{figure:papov}
  \end{figure}

Moreover, we `clip' this library for the highest IR
luminosities, using the \LIR{}=$10^{12.2}$~\Lsun{} SED shape for
galaxies more luminous than this value. This makes the SEDs colder
than in CE01 for ULIRGs and makes it compatible with most of the known
data, from low-redshift, low-luminosity to high-redshift luminous
galaxies \citep[see CE01,][]{PapRudLeF07}. Figure~\ref{figure:papov}
presents the relation between 24\um{} and total IR
luminosities that is predicted by the ``clipped'' CE01 library, together with
observations from \citep{PapRudLeF07} in the range $z=1.5-2.5$.
This clipping can be justified by the SEDs of high
luminosity galaxies (ULIRGs) like those seen at high redshift being
poorly known in the local universe: their dust temperature can only
be measured by Herschel. Therefore, the CE01 SEDs are extrapolated
within this luminosity range. We chose to clip the IR dust temperatures of
the CE01 library to those of the luminosity range really observed in
the local universe, rather than extrapolating them.

\section{Results}
\label{section:results}

In this section, we present the counts, luminosity functions, and
SF history that are modeled from the counts inversion, as
well as the effect of using or not using low-z luminosity functions as priors. We
compare these results and their redshift decompositions to
measurements obtained from direct methods and bibliographic data only in
Sect.~\ref{section:validation}.

\subsection{Counts inversion}

\begin{figure*}[!htbf]
  \includegraphics[width=0.98\textwidth]{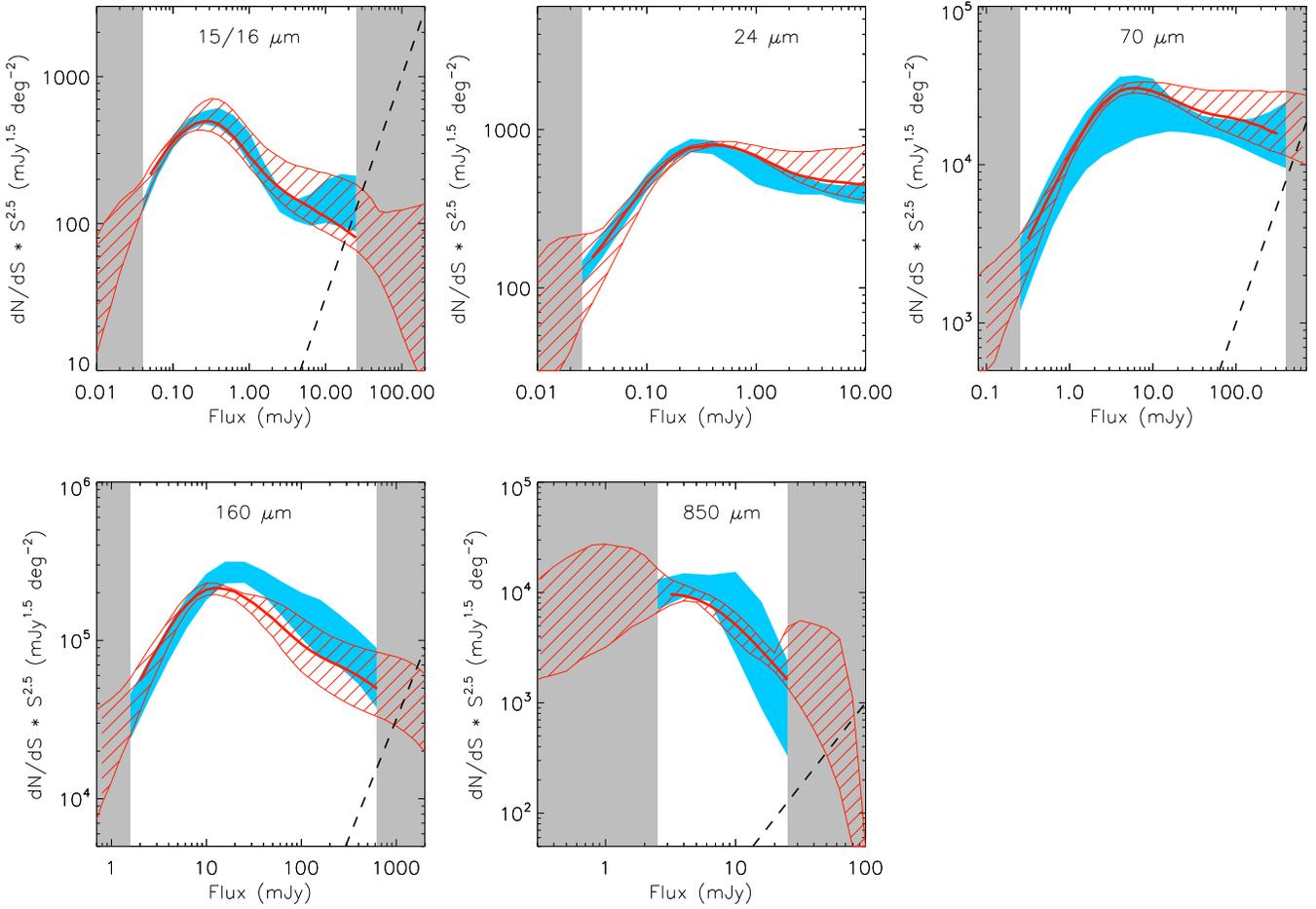}
 \caption{Mid- and Far-IR counts obtained from the non-parametric inversion of the
 observed counts. The data counts that are the basis of the
 inversion are represented by the shaded blue zones (fitted values $\pm$ 1
 $\sigma$, corresponding to the
 zones of Fig.~\ref{figure:obsdata}). The best fit to the counts
 is shown by the solid red lines, together with the range of allowed
 solutions (red dashed areas).  The vertical gray zones delimit the flux
 ranges where some counts are observed. All the counts
 modeled here satisfy the CIRB constraints. The oblique dashed
 line corresponds to a statistical limit of a one square degree survey (less
 than 2 sources per decade in flux).}
 \label{figure:countsok}
\end{figure*}

This inversion model has been designed to reproduce infrared galaxy
counts and as a result to derive a range of possible total IR LF as a
function of redshift, that can be converted afterwards into a range
of cosmic SF histories. The success of the model can be visually tested by
comparing the range of predicted galaxy counts with the observed
number counts and their dispersion (see Fig.~\ref{figure:countsok}).

At first glance, one can see that the observed counts are well-fitted over the
whole IR range, from 15 to 850\,$\mu$m. In particular, the bumps at 15
and 24\,$\mu$m are reproduced simultaneously. The 850\,$\mu$m differential
counts are well-fitted too: negative k-corrections make it possible to
see a high-redshift population of galaxies, namely ULIRGS and HLIRGS
(\LIR{}$>10^{12}$L$_\odot$) at $z>2$, which are hardly seen at other
wavelengths except in faint 24 and 70\,$\mu$m counts.  

Interestingly, one can see from Fig.~\ref{figure:countsok} that, while
the only strong constraint that is imposed on the model is to keep a smooth
dependence of the LF with redshift and luminosity, the model is unable
to perfectly fit the observed number counts and their dispersion at
all flux densities and wavelengths even though the whole range of
possible LF and associated redshift evolution has been spanned
blindly.  Some solutions tend not to fit the 15\,$\mu$m
counts perfectly, the 70\,$\mu$m counts are slightly overproduced, and more
strikingly, the 160\,$\mu$m counts are underproduced around 20~mJy. 

A major strength of this model is to provide an
objective and statistically significant way to test a given library of
template SED.  Indeed, the discrepancy between the model and observed counts
cannot arise from the LF itself since it was allowed to vary both with
luminosity and redshift with a high degree of freedom \citep[see
also][for a discussion on the corresponding biases in a slightly
different context]{OcvPicLan06b}. It must therefore arise from the library
of template SEDs that is used as an input for fitting the number counts
(through the ``k+d'' matrix $\mathbf M$). The CE01 library of template
SEDs that is used here reflects the median trend of local galaxies and
was found to be statistically consistent with the radio-infrared
correlation up to $z\sim$1.3 \citep{ElbCesFad99,AppFadMar04}, with the mid-infrared observations of galaxies up to
$z\sim$1 with moderate variations \citep{MarElbCha06} and with massive galaxies selected with the BzK technique \citep{DadDicMor07}. The origin of the discrepancy can therefore come from
three possible origins. The first two possibilities compatible with no
evolution of the infrared SED of galaxies are (i) a
bias towards cold galaxies due to the shallow depth at 160\,$\mu$m
within the dispersion already existing at $z\sim$0; (ii) the existence
of a subpopulation of cold galaxies, already present locally but not
yet identified due to limited constraints on both sides of the peak
emission in the far infrared. A third possibility would be that the
infrared SED of galaxies evolve as a function of redshift \citep[see e.g.][]{ChaSmaIvi02}.

It is not possible to disentangle between the three possibilities
based on the present dataset. However, a forthcoming paper will study
this issue in detail using a stacking analysis at 160\,$\mu$m
(Magnelli et al., in preparation).  It must be noted though that the
method described here is versatile enough to allow its user to test
any library of template SED against existing constraints from galaxy
counts and the infrared background. It will therefore be
a straightforward matter to check whether any change or evolution in the SEDs can reproduce
the number counts at all wavelengths and flux densities. We refrained
from making these adjustments to the counts at this stage since any of
the previously mentioned alternatives is equally possible. 


\subsection{Solutions: Range of evolving luminosity functions}

\begin{figure*}[!htbf]
  \includegraphics[width=0.98\textwidth]{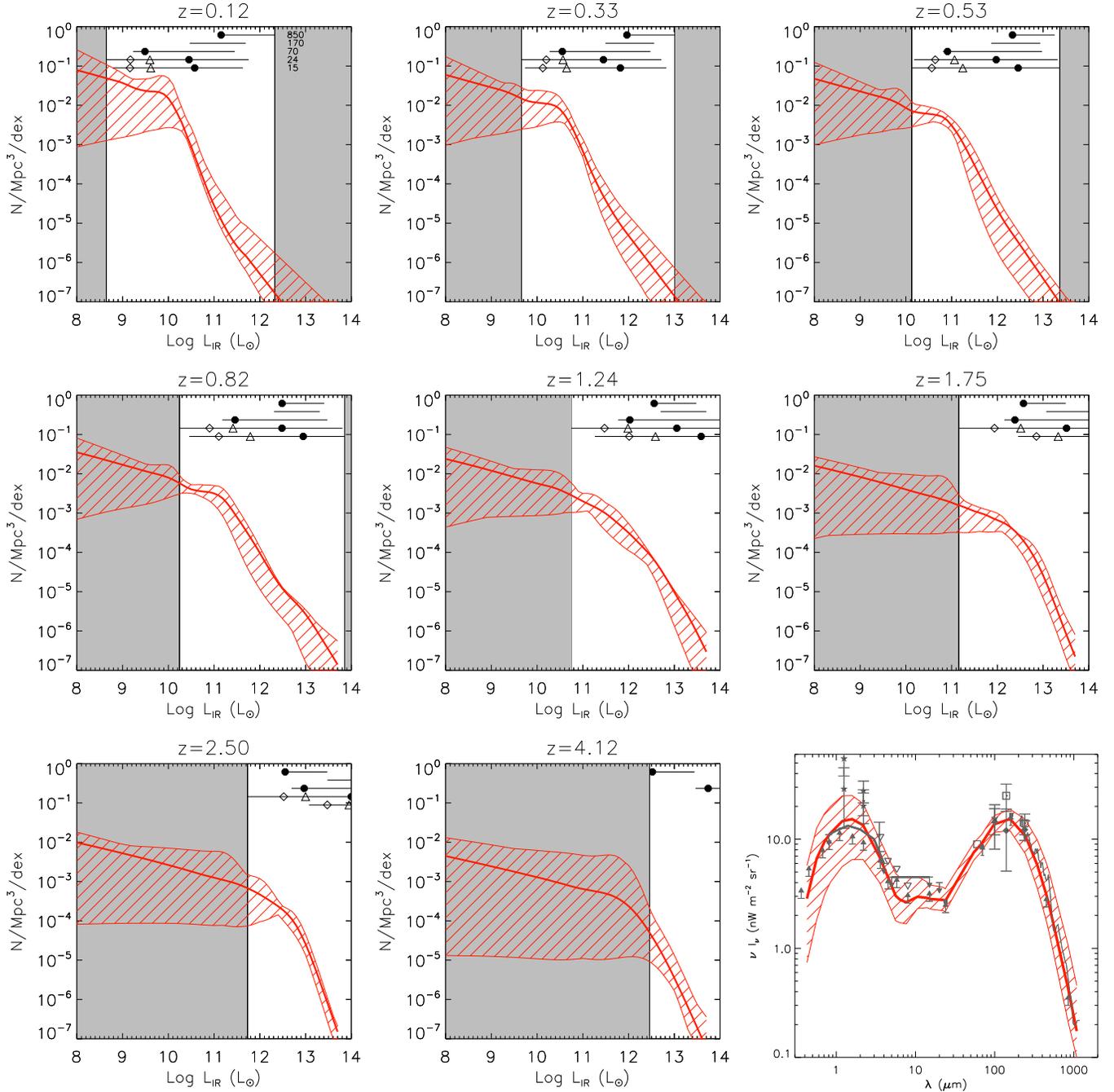}
  \caption{All possible solutions for the evolving LFs that best reproduce
  all the IR counts of Fig.~\ref{figure:countsok} and the CIRB
  constraints. As in Fig.~\ref{figure:countsok}, the thick red line and
  the dashed red areas represent the best-fitting solution and the range of
  allowed solutions, respectively. By construction, no conclusion on the
  LF can be made in the vertical gray-shaded areas where a divergence is
  expected because these objects are not seen in the counts because of the flux limits of
  current IR surveys. At the top of each panel, the ranges of \LIR{}
  probed by IR counts are shown as horizontal lines, from 15 to 850\um{}
  (bottom to top). Diamonds, triangles, and filled circles correspond to
  0.1, 0.3, and 3~mJy fluxes, respectively.  Bottom right panel: range of
  CIRB corresponding to the inverted LFs. Data points (in gray) are
  taken from the compilation by \citet{DolLagPug06}.}
  \label{figure:lfsok}
\end{figure*}

Looking now at the main output of the model, i.e. the evolution of the
total IR luminosity function with redshift, we note that large parts
of the LF are constrained very well by the inversion (see Fig.~\ref{figure:lfsok}). In
particular, the number density of galaxies with $10.5 < \log_{10}
L_\mathrm{IR} < 11.5$ at $z<0.5$ is tightly constrained, as are
the numbers of galaxies with higher luminosities at higher redshift.

We will present in Sect. \ref{section:lfir} a comparison of these
solutions to the  \LFIR{}s obtained from direct measurements at low redshift. 

But before doing so, we note that a fraction of the solutions present a knee in
the LF, particularly at $0.3<z<1$ around $10^{10}-10^{11} L_\odot$, which
is partly responsible for the bump seen in the number counts at 24 and
15\,$\mu$m at the corresponding flux densities. However, not all
solutions of the inversion technique present such a strong knee, which
might be seen as an artifact.

\subsection{Evolution of the star-formation activity}
\label{section:SFRD}

\begin{figure*}[!tbf]
\includegraphics[width=0.99\textwidth]{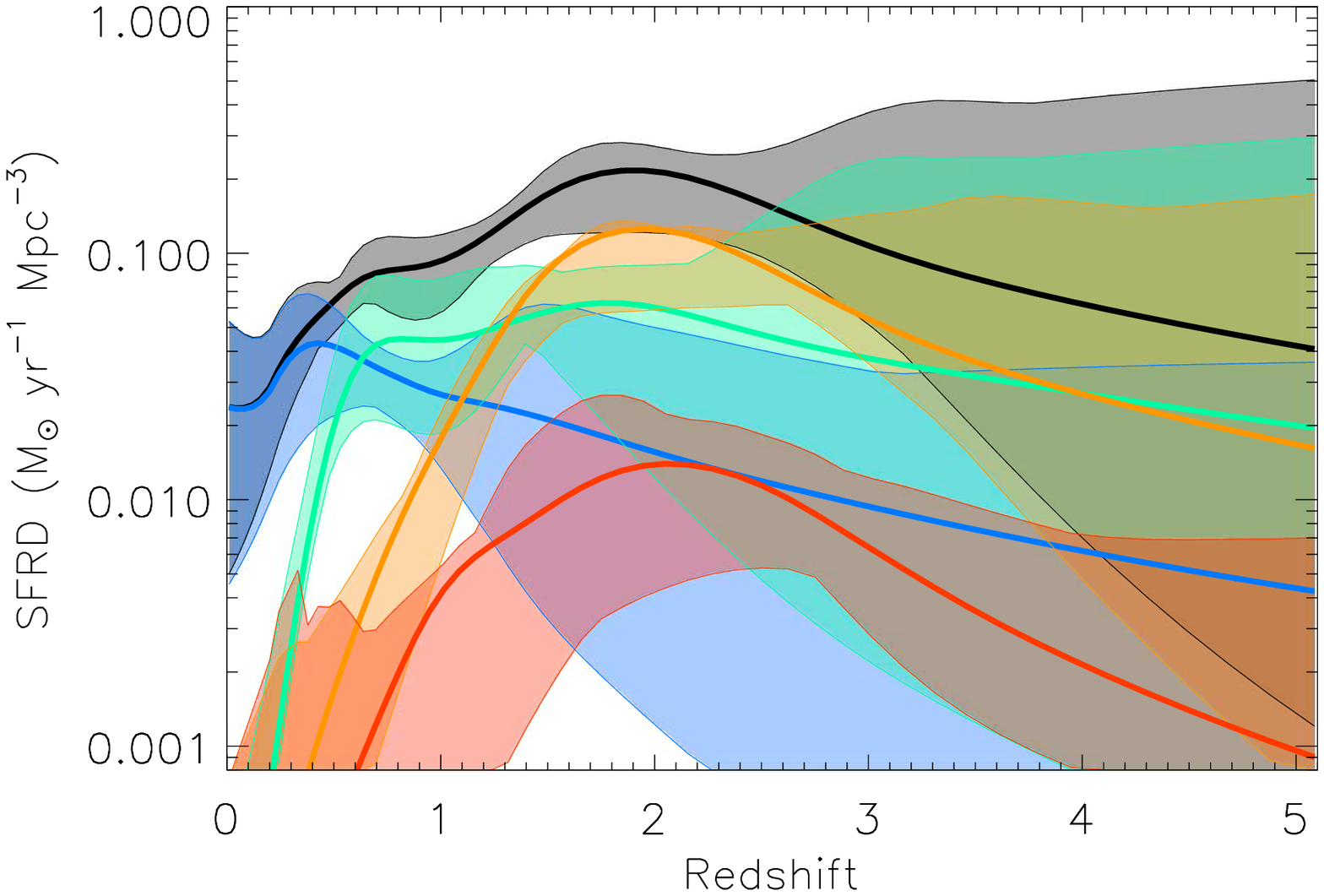}
\caption{Star-formation rate density since $z=5$ derived from the
  counts inversions. The SFRD is obtained from the range of all possible
  luminosity functions derived from the counts and respecting the CIRB
  constraints. The solid lines correspond to the best fit to the
  counts and the transparent shaded areas show the range of uncertainties.
  From top to bottom at z=0.8: black=all galaxies, green=LIRGS,
  blue=normal star-forming galaxies, orange=ULIRGS, red=HLIRGS.}
\label{figure:sfrdmodel} 
\end{figure*}

The SFRD can now be estimated from the LFs that were obtained from the counts
inversion. The total infrared LFs are integrated over the whole range of luminosities
down to $10^7$~L$_\odot$. The resulting total infrared luminosity density
is then converted into a SFRD using formula (\ref{EQ:kenn})
\citep{Ken98}:
\begin{equation}
{\rm SFR ~[M}_{\odot}~ {\rm yr}^{-1}]=1.72 \times 10^{-10} ~{\rm L}_{\rm
 IR} ~[{\rm L}_{\odot}] \,.
\label{EQ:kenn}
\end{equation}
Therefore, the regions shown in Fig.~\ref{figure:sfrdmodel}
effectively represent the range of \emph{all} possible SF histories
that are compatible with the multi-$\lambda$ counts and the CIRB. The
SFRD history obtained from the \LFIR{} that produces the best fit to
the counts is also shown for each luminosity class of galaxies.

This inversion shows that, indeed, an IR downsizing is at work:
``normal'' galaxies dominate the SFRD at low redshift (although the
contribution of ULIRGS and HLIRGS are poorly constrained in the
low-$z$ range because they would correspond to bright and very rare
sources, not easily seen in the counts of current deep surveys). At $z>0.8$, LIRGS
dominate the SFRD, whereas the contribution of ULIRGS peaks at $z \gtrsim
2$.
These results will be compared in detail to measurements from direct methods and
bibliographic data in Sect.~\ref{section:validation_SFRD}.

As mentioned before, it is also possible to use an additional prior
for the inversion: direct measurements of the 8 or 15\um{} LF below
$z=2$. Although subject to many caveats (e.g. the strong dependency
on the PAH modeling of the SEDs), they can be used as a prior to guide the
inversion, and at least constrain the solutions at
low redshift\footnote{Technically, this supplementary prior involves
  again penalizing  the formal $\chi^2$ by adding an extra term that
  measures the distance between the observed LF and the solution (this
  corresponds to $\mu\neq 0$ in Appendix~\ref{s:NP}). The LFs that
  are too different from the known low-redshift LF are therefore
  strongly penalized, so excluded {\it de facto}.}.

We checked the effect of using this prior from direct measurements at
low redshift. We observe that some uncertainties in the LIRGS and
ULIRGS contributions are slightly tightened, but we also note that
the trends are basically unchanged. We interpret this surprising result as
follows: the leverage that we have access to by inverting galaxy
counts on a very wide wavelength-basis (from 15 to 850\um{}) is large
enough to provide a realistic description of the redshift distribution
of the sources on a statistical basis. This is likely to only be possible
because the library of SEDs that we use seems close to the real SEDs
of galaxies, on average (again, in the statistical sense) at any
redshift lower than $z\simeq3$, thus avoiding a complete blurring of
the de-projection of multi-wavelength galaxy counts onto the
luminosity function space.  Therefore, various populations of galaxies
at different redshifts are seen at various wavelengths, which 
considerably reduces degeneracies and enables us to recover the history of IR
galaxies as a whole.

Since the philosophy of this paper is to remain as conservative as
possible, we choose not to use the low-redshift LF measured from
direct methods as a prior in the following. Indeed, doing so would only slightly
change our results, and it would introduce a source of potential
additional errors propagating from the errors intrinsic to direct
methods (k-corrections in the mid-infrared or redshifts).

\section{Validation: comparison with direct measurements at low redshift}
\label{section:validation}
After presenting the global outputs of the model in the previous
section, we detail here the redshift decomposition of the inversion
results and we compare them to measurements obtained from direct
(redshift-based) methods. This comparison is particularly challenging
since no redshift information was used as an input in the inversion.

\subsection{Redshift decomposition of the mid-IR counts}
\label{section:validation_counts}

\begin{figure*}
  \includegraphics[width=0.98\textwidth]{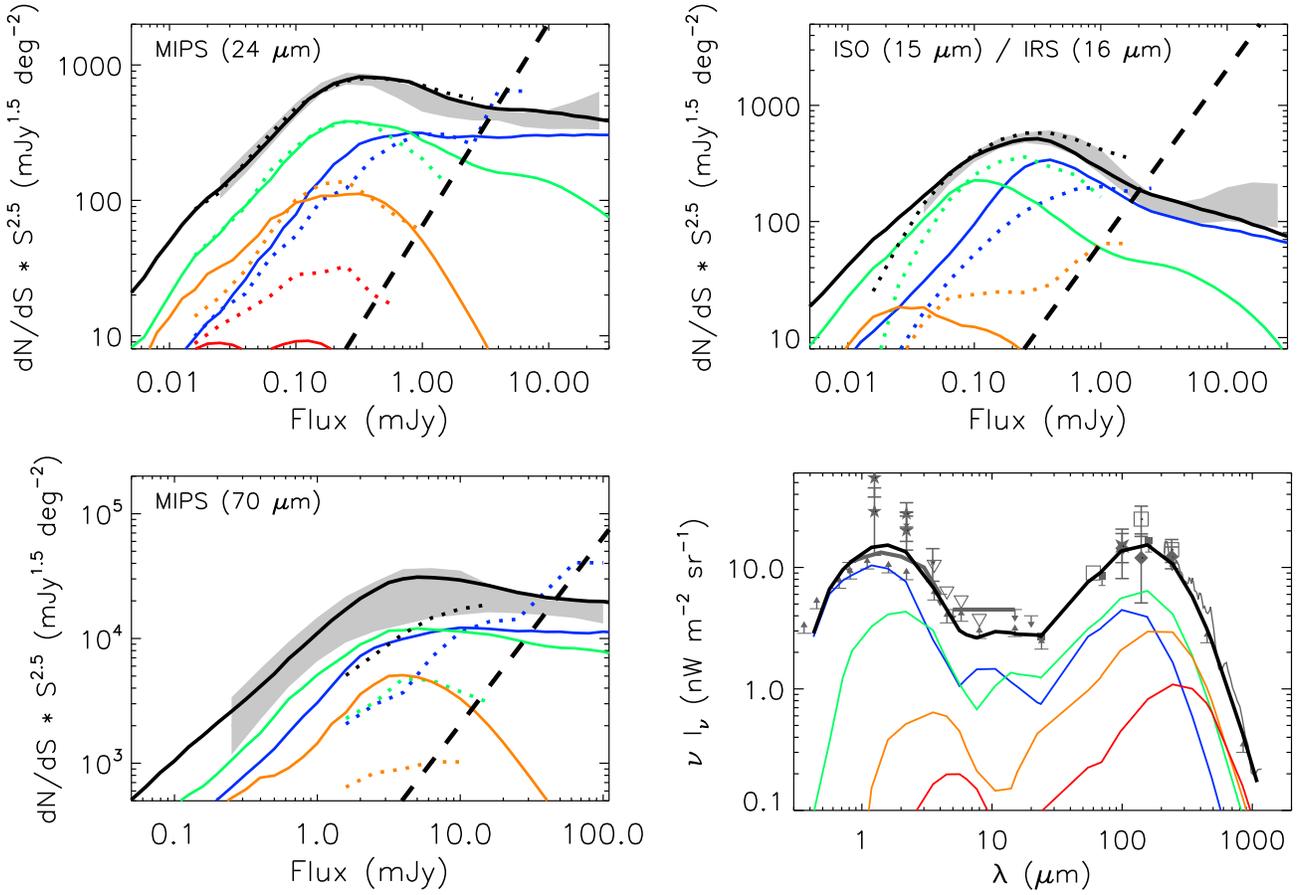}
  \caption{{\sl Top and bottom left}: Counts at 16, 24, and 70\um{} modeled (solid
 lines) from the inversion, on the one hand, and observed (dotted lines)
 in GOODS fields, on the other. The counts are
    decomposed in redshift bins (black=all redshifts, blue=$z<0.5$, green=$0.5<z<1.5$,
    orange=$1.5<z<2.5$, red=$z>2.5$). {\sl Bottom right}: CIRB decomposed in
    redshift (same color-coding). The oblique dashed line corresponds to
    the limit in statistics due to the smallness of a field like GOODS
    North+South or 0.07 square degrees: less than 2 galaxies per
    flux bin of width $\delta \log$F=0.1~dex are expected below
    this limit.}
  \label{figure:zdists_spitzer}
\end{figure*}

To validate our inversion results, we need to compare the redshift
decomposition of the IR counts to observations from a direct method.
To do so, we compared our results to data from the GOODS survey
(P.I. Dickinson for GOODS-Spitzer, P.I. M.Giavalisco for GOODS-HST)
originally presented in \citet{GiaFerKoe04}. This survey consists of two
fields which have been subject to several studies at various
wavelengths in the past few years.  We investigated the redshift
decomposition of the counts at 16, 24, and  70\um{} using the optical
counterparts of these sources in GOODS and making use of the
spectroscopic and photometric redshifts.  The sample that we used
covers a total area of 0.07 square degree on the sky. Although this
area is quite small and cosmic variance might affect our study, we
found that the luminosity functions measured from the direct method are
similar in both GOODS fields, making them compatible within
2$\sigma$. The spectroscopic completeness is high (60\% at $S_{24\mu
m}=30$\uJy{} for $z<1.5$ sources), and we complemented them with
photometric redshifts computed with the code Z-Peg \citep{LeBRoc02}
with a precision $\Delta z/(1+z) \simeq 0.04$ to 0.1 depending on the
redshift of the sources.  We then used the Vmax formalism to correct
from incompleteness at the lower flux limit. The galaxies showing
signs of AGNs (identified from X-rays or optical emission lines), were
excluded from the sample of 24\um{} sources. Doing so enabled us to
use SEDs of galaxies to compute k-corrections and only
slightly affects our results, mainly at the very high-luminosity end of the
luminosity function at moderate ($z=1$) or high ($z>2$) redshifts.

Figure~\ref{figure:zdists_spitzer} presents the resulting comparison of the
redshift decomposition of the counts obtained from direct and inverse
methods. At 24\um{}, our best solution for the recovered \LFIR{} indeed
produces a redshift decomposition of the counts that is compatible
with the observed ones. The match of 15\um{} counts as a function of
redshift is poorer because CE01 templates represent the
fluxes at this wavelength less well for $0.5<z<1$ galaxies
\citep[e.g.][]{MarElbCha06}. As for the 70\um{} counts, the observed
decomposition is not complete at faint fluxes, making the comparison
hazardous.

\subsection{Comparison to direct measurements of the infrared luminosity functions}
\label{section:lfir}

\begin{figure*}[!tbf]
    \includegraphics[width=0.98\textwidth]{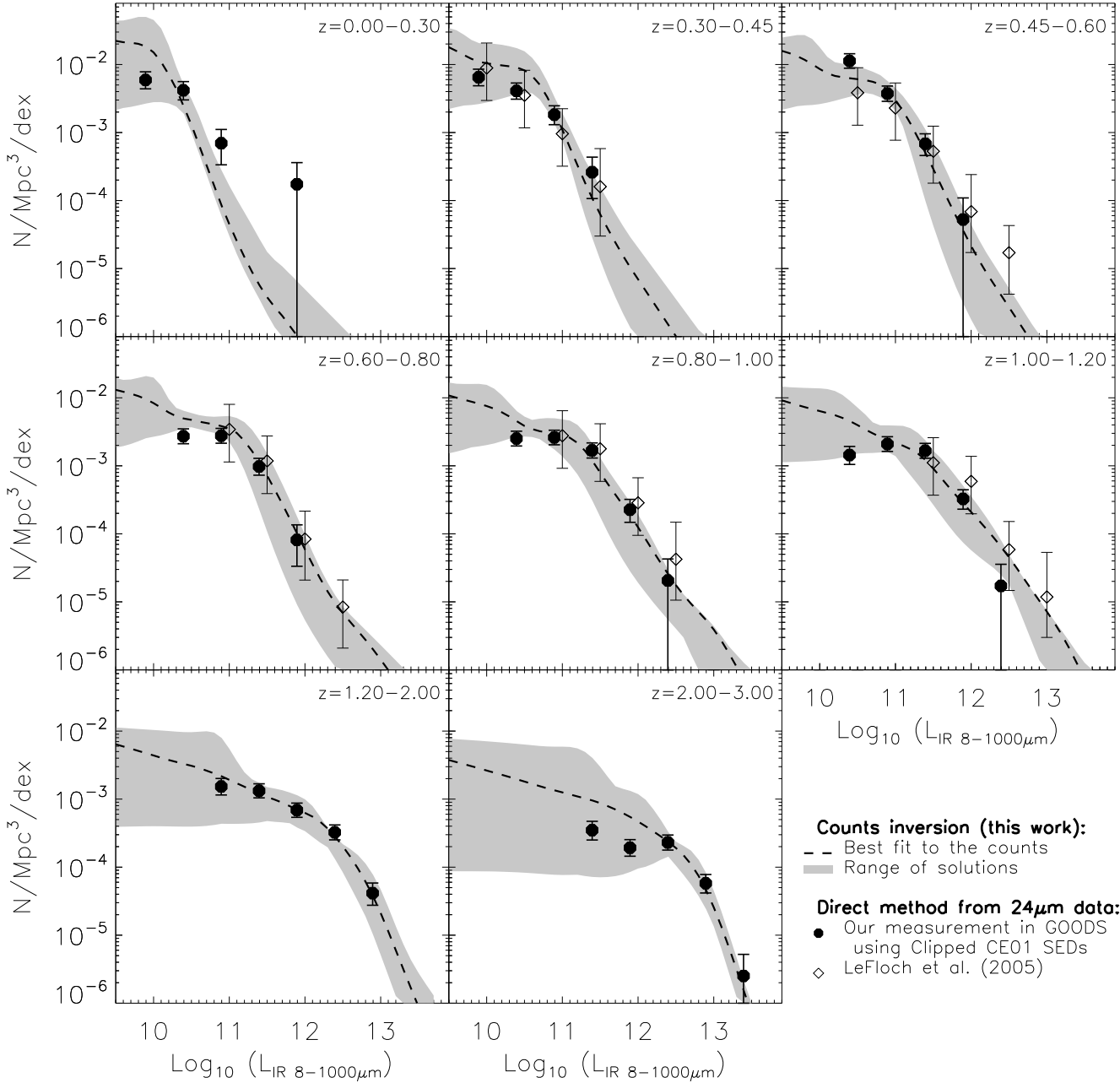}   
    \caption{Total IR luminosity functions derived independently from the counts
   inversion and from direct measurements. ``Clipped'' CE01 templates are used for the 24\um{} fluxes
    to \LIR{} conversion (filled black dots). The measurements from \citet{LeFPapDol05} are also
   shown for reference. The
    model (counts inversion) curves (dashed lines) are given at the mean redshift
    of each bin. The shaded areas represent the range of uncertainties
    on our modeled LF at the mean redshift of each bin.}
    \label{figure:l15} 
  \end{figure*}

We now compare the range of luminosity functions \LFIR{} obtained from our inversion to
some measurements of the \LFIR{} obtained from the direct method.
As noted before, several studies \citep{LeFPapDol05,BabRowVac06,CapLagYan07}
have measured the 15 or 8\um{} luminosity functions, which can be
converted to \LFIR{} if a library of SEDs is assumed.
We show in Fig.~\ref{figure:l15} the results from \citep{LeFPapDol05}
for reference. 
We also provide on the same figure our own direct measurements of the evolving
\LFIR{} that we derived from 24\um{} galaxies seen in the GOODS fields
(North+South). 
We chose to do these direct measurements again for two
reasons. First, for consistency with our inversion, we used the same
library of templates for the k-corrections (used for MIR to total LIR
conversion)\footnote{Note that \citet{LeFPapDol05} errors bars are rather
large because they  partly reflect the uncertainties in the SEDs of
galaxies. We checked that
using another library of SEDs (e.g. the LDP03 library) produces
differences in our LFs comparable to the error bars of
\citet{LeFPapDol05}.}. Second, our data at 24\um{} reaches a depth of 24\uJy{}
instead of 80\uJy{} in previous works, extending the LF to the
faint end. The completeness at this depth is 85\% \citep{Cha07}.
Not surprisingly, our direct measurements of the \LFIR{} from 24\um{}
fluxes are compatible with previous studies that used the same
method. In particular, we find the same values as \citet{LeFPapDol05}
who explored the $z<1$ domain.  However, we reach higher redshift
galaxies thanks to the depth of our sample. However, k-corrections are very
uncertain for these high-$z$ sources (24\um{} corresponds to less than
8\um{} at $z>2$, a range where the library of SEDs is not
validated). Therefore, we use these high-redshift measurements with
caution, as mentioned earlier.

It is striking in Fig.~\ref{figure:l15} that the LF obtained from the direct method is
consistent with the best-fitting LF derived from the counts inversion
in the common range that
they probe. This is remarkable
because no information on the redshift of the sources was used in the
inversion.  This means that all the constraints that one can get from
these direct measurements are not really needed for the inversion:
although we do not use the redshift of the sources, we recover the
observed redshift distributions that are here expressed equivalently
in terms of evolving luminosity functions. The interpretation is the
same as for the redshift decomposition of the counts, which are 
different views of the same phenomenon.

The meaningfulness of the agreement between the observed and the
recovered LFs is strengthened by the fact that that both \LIR{}
luminosity functions are measured or estimated using the same library
of SEDs, thus using the same k-correction in a consistent way. Using
another library of SEDs for both methods would produce the same kind
of agreement, although the precise shape of the LFs would be slightly
different from what we obtain here with the ``clipped'' CE01 library.

\subsection{Cosmic star-formation history vs literature}
\label{section:validation_SFRD}

In this section, we compare the SFRD history that we derived from the
multi-$\lambda$ counts inversion to ``direct'' measurements at low
redshift.

\subsubsection{Total SFRD}

\begin{figure*}[!tbf]
\includegraphics[width=0.89\textwidth]{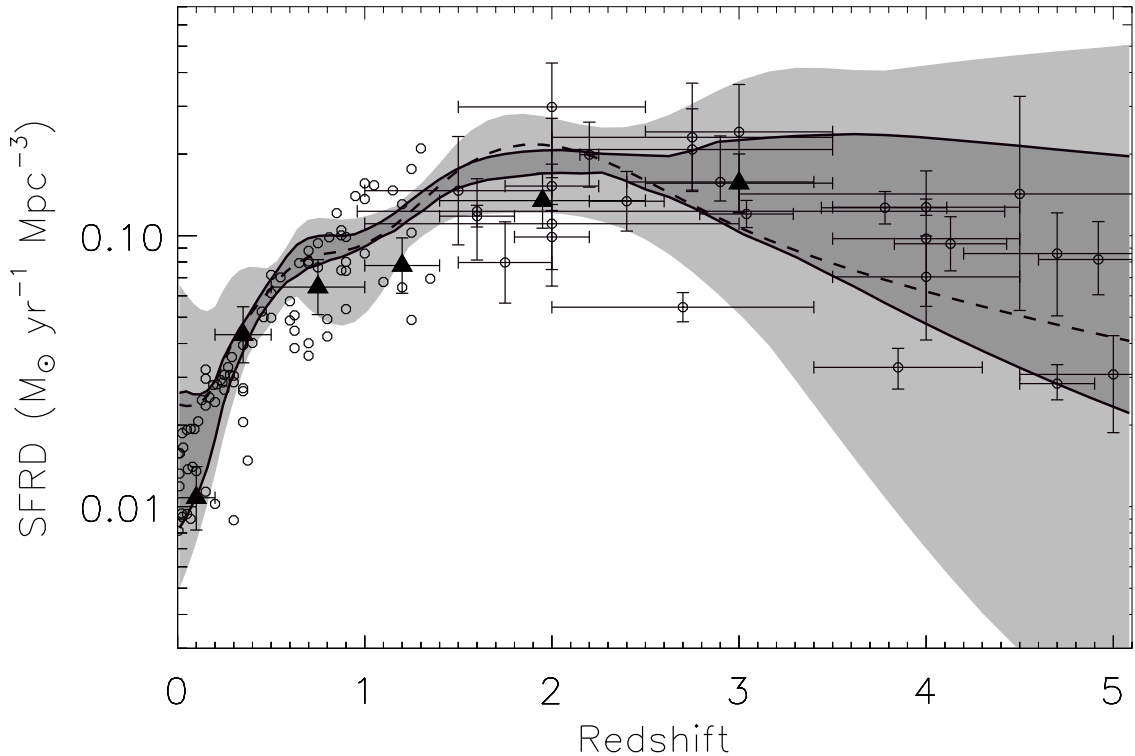}
\caption{Total SFRD regions compared to the compilation of direct
  measurements from \citet{HopBea06} (empty circles and error bars
  only for $z>1.5$ for clarity purposes). The outer light-gray regions
  corresponds to all possible LFs fitting the counts. The inner darker
  region includes 68\% of these models. The dashed line corresponds to
  the LF producing the best-fit to the multi-$\lambda$ IR counts. 
    The
  solid triangles with error bars show the integrated SFRD that we measured
  in the GOODS fields in several redshift bins (see also
  Fig.~\ref{figure:sfrdobs} for a luminosity decomposition of these measurements).}
\label{figure:sfrdmodel_and_hopkins} 
\end{figure*}

The comparison of the total SFRD derived with this inversion technique
matches the compilation of direct measurements from
\citet{HopBea06} (Fig.~\ref{figure:sfrdmodel_and_hopkins}).
Moreover, we also show in Fig.~\ref{figure:sfrdmodel_and_hopkins} the
SFRD that we derived from 24\um{} data in the GOODS fields. We must stress here that these measurements, obtained with
the ``direct'' method, are independent of the SFRD inferred by the
counts inversion\footnote{The datasets are independent, indeed. But
strictly speaking, it is actually not the case for the SFRD measurements
because the libraries of SEDs used for k-corrections are the same in
both methods.}. We only present them here for comparison to the
inversion results in this subsection. The total infrared luminosities measured for
individual 24\um{} sources in the GOODS fields were summed up in redshift bins
and converted to SFRDs.
Like several authors before us
\citep[e.g.][]{FloHamThu99,LeFPapDol05,PerRieEga05}, we could have
extrapolated the measured luminosity functions in the
faint end to obtain the total SFRD. But this is somewhat dangerous and depends strongly on the
parameterization of the LF fits, so we chose not to
extrapolate the luminosity functions at faint fluxes to estimate the
SFRD. Instead, we used only the sources brighter than our flux limit at
24\um{} (24\uJy{}) to estimate lower-limits for the SFRD at every
redshift. Therefore, all the points in this figure should be
considered as lower limits.

Our inversion technique, on the other hand, makes it possible to
partially avoid such caveats. First, unlike direct methods, more than
one band is used to estimate the total IR LF. Second, the
extrapolation of the LF in the faint-end is achieved automatically via
the only constraint of having a smooth variation in \LIR{} and
$z$. Therefore, the area of uncertainties we propose here are very likely
more robust than previous estimates because they use more data and are
less dependent on parameterization for both the shape of the LF and
its evolution.

The SFRD obtained from the counts inversion and our measurements with
the ``direct'' method in GOODS are in rather good agreement, which tends
to give credit to the inversion. One might note, however, that the data
points are systematically lower than the inversion results, which
illustrates the choice of not extrapolating the measured LFs at faint
luminosities to estimate the SFRD.

\subsubsection{Luminosity decomposition of the SFRD}

\begin{figure*}[!tbf]
\includegraphics[width=0.89\textwidth]{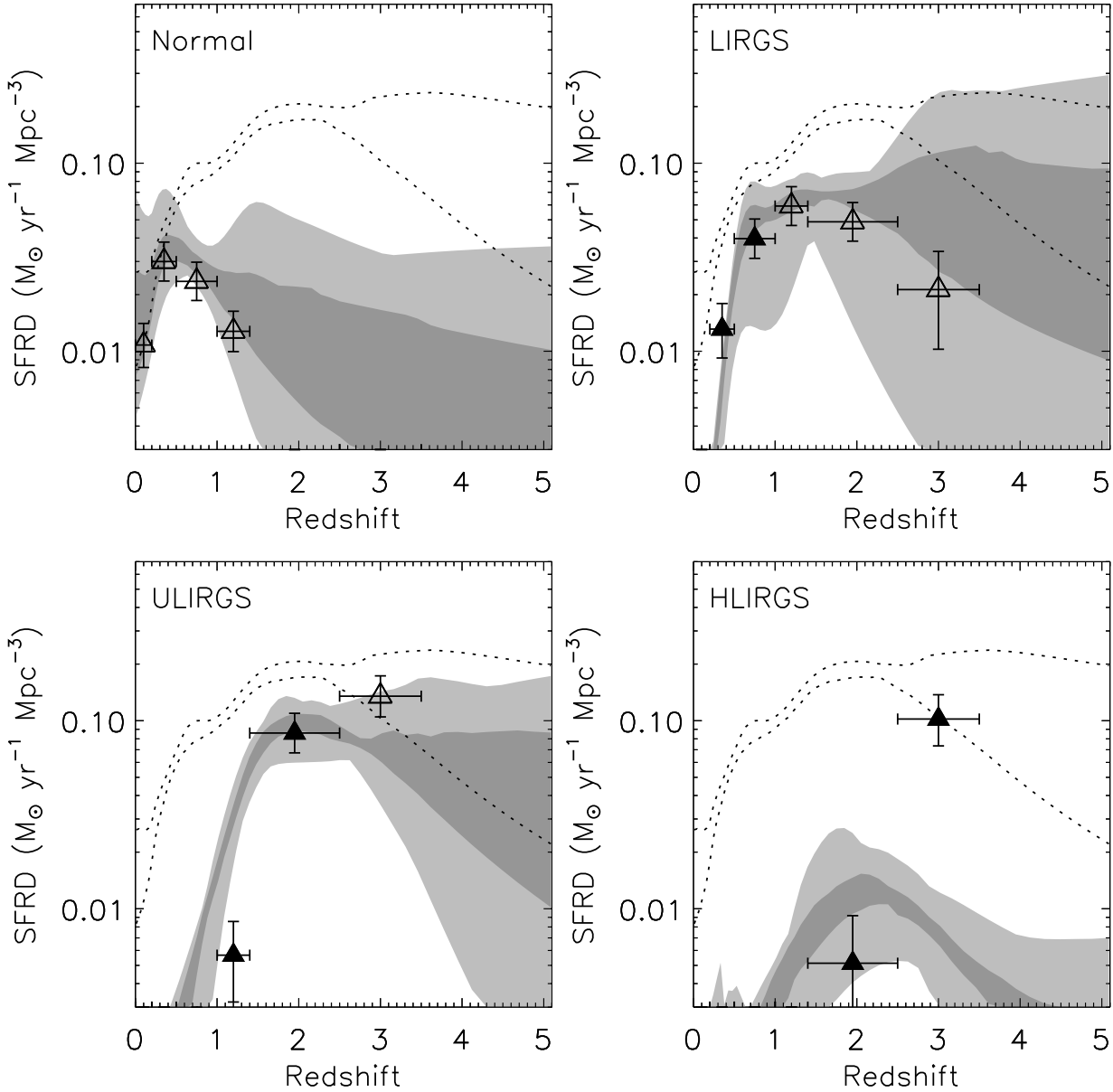}
\caption{History of the SFRD decomposed in four
 infrared luminosity classes. The inversion results (full set of models
 and 68\% inner region) are shown with gray-shaded areas. The total SFRD
 history (68\% inner region) is also shown for reference in each panel
 as dotted lines. Our direct (and independent) measurements from the GOODS
 survey are shown as triangles. Empty triangles are used for bins
 affected by completeness. The $z=3$ point for HLIRGS is subject to
 caveats (see text for details) so is probably overestimated.}
\label{figure:sfrdobs} 
\end{figure*}

Figure~\ref{figure:sfrdobs} presents a detailed comparison of the
history of the SFRD inferred from our
inversion and decomposed in luminosity classes to what can be
independently obtained from the ``direct'' method with 24\um{} sources
in GOODS.

Let us first consider the reliability of the direct measurements used
for comparison.
Because we chose not to extrapolate the LFs at low luminosity, several points in this figure must be
considered as lower limits. It is not the case, however, for the $z<1.2$ points for LIRGS and
the $1<z<2.5$ points for ULIRGS. Overall, we are limited at very low redshift
by small statistics and at high redshift by flux limits.
Moreover, the contribution of HLIRGS should be taken with caution. Indeed, by
inspecting these luminous high-redshift sources, we came to the
conclusion that a large fraction of the photometric redshifts computed
for these IR-bright sources have poor precision, leading to
catastrophic failures for almost half of the HLIRG sources at
$z>1.5$. This poor performance of the photometric redshifts is not too
surprising for this class of galaxies at such high redshifts since the
templates used in the fitting procedure have a relatively low level of
dust, compatible with most galaxies seen in the current optical and
NIR surveys.
These direct measurements present a nice picture of the ``IR
downsizing'', where the
cosmic SFRD was dominated by brighter and brighter galaxies in the IR
as we go back in time. Our results confirms the IR view of the
cosmic SF history that was explored in previous works up
to $z=2$ \citet[e.g.][]{CapLagYan07}. At $z=2$, we confirm that ULIRGS
seem to dominate  the budget of the SF activity.

A number of interesting remarks arise from the comparison of the
``inverted'' SFRD to the ``direct'' measurements.
First, it is comforting to see that both methods give consistent
views of the IR downsizing. In both cases, low luminosity galaxies
dominate the SFRD at $z<0.5$, and ULIRGS are dominant at $z>2$.
However, a more detailed comparison of both results provides
interesting clues to what is really seen in current deep surveys.  One can easily notice that most data points (``direct''
measurements) are at the lower limit of the area allowed by the
inversion. This means that the 24\uJy{} limited sample of 24\um{}
galaxies probe almost all the cosmic SFRD. However,
the same remark also opens up the possibility that up to 50\% of the
SFRD is not yet resolved in sources down to our flux limit, especially
for low-luminosity galaxies at $z>0.5$. 

Another interesting point is that the inversion does not allow the
population of HLIRGS to contribute much to the CIRB at \emph{any}
redshift. This somewhat contradicts the direct measurements for the
same objects, which tend to indicate an increasing contribution of
these extreme sources at $z>2$. But again, we must stress that many
uncertainties lie in the observations of these distant sources
(photometric redshifts, validity of the SEDs, contribution of AGNs,
etc.). Therefore, we must conclude that this population cannot be too
numerous to reproduce the deep IR counts, including the 850\um{} ones,
if the ``clipped'' CE01 SEDs are used at any redshift. We checked that
allowing more HLIRGS at these redshifts, at a level comparable to the
observed number, overproduces the 850\um{} counts. This means that
either these objects do not exist (and indeed, as we mentioned before,
about half of these HLIRGS have a wrong photometric redshift, hence a
wrong luminosity), or the SEDs for these objects are very different
from the templates in the SED library that we use. Of course, both
reasons may be at work simultaneously.  Interestingly, if the original
CE01 library is used (both for the inversion and for direct
measurements), the situation is similar: the SFRD of HLIRGS inferred
by the inversion of the multi-wavelength counts is still smaller (by a
factor of 4) than the value inferred from direct measurements of
24\um{} sources. The main difference, in this case, is that the SFRD
of HLIRGS are roughly a factor of 5 to 10 larger than what is derived
with the clipped CE01 library.

Finally, we notice that the measured SFRD for ULIRGS  at $z<1.5$ (see
Fig.~\ref{figure:sfrdobs}) is smaller than the range allowed by the
inversion. This could be explained by cosmic variance due
to the small area covered by our GOODS sample.

\subsection{Evolution of the stellar mass density}
\label{section:mstar}

In this section, we address the question of the consistency of
the SFR history that we derive from the inversion
model with the independent observational constraints existing on its
integral, namely the evolution of the comoving
density of stars per unit comoving volume. After assuming an
initial mass function (IMF) and computing
the mass of stellar remnants after the death of massive stars, it is
straightforward to compute the total amount of stars that must be
locked into galaxies as a function of redshift, on the basis of the
SFR history. In our computation of stellar
masses, we account for the recycling of stellar material into the ISM,
for the mass of stellar remnants (which account for about 15\% of the
total stellar mass at $z=0$) and for the evolution of the metallicity,
using the spectral synthesis code P\'EGASE.2
\citep{FioRoc97,FioRoc99}.

This is not the first time that such a computation has been performed, but we
believe that this is an important test that has been the source of
discussions in the recent past, in particular with the claim that both
histories - SF and stellar masses - were not consistent
with the integral of the SF history producing more
stars than actually observed at any redshift. Our paper now proves
this claim to be incorrect.
   
Before discussing our own computation, we wish to emphasize an important
point regarding the effect of the choice of a particular IMF in this
process. Although various IMFs have been proposed in the past, including
top-heavy IMFs for starbursting galaxies \citep[see
e.g.][]{ElbArnVan95,ElbArnCas92,RieLokRie93,LacBauFre08,Dav08}, no
definitive evidence has been provided yet for a non-universality of the
IMF. The main difference that is now commonly accepted with respect to
the pioneering work of \citet{Sal55} is the finding that the slope of
the IMF changes around 1 M$_{\odot}$, in the direction of having a lower
contribution of low mass stars to the total mass of stars formed, or
equivalently a larger contribution of stars more massive than 1
M$_{\odot}$ \citep[see the review by][]{Cha03}. Nonetheless, such a change in the IMF
almost equivalently affects both the conversion factor used to determine
the SFR from \LFIR{} and the mass-to-light ratio used to derive the stellar
mass. In the present study, the SFR is derived from the total IR
luminosity assuming the coefficient computed for a Salpeter IMF by
Kennicutt (1998, see Eq.~\ref{EQ:kenn}). For a different IMF, such as
the Baldry \& Glazebrook (2003, hereafter BG03) one which shows a
flattening below 1 M$_{\odot}$ as discussed above, the SFR would be 0.45
lower and the M/L ratio would also be reduced by a similar factor (0.6,
computed using the PEGASE.2 code).

The evolution of the cosmic stellar mass density with redshift that we
computed by integrating the SFR history resulting from the inversion
model is found to be in good agreement with the latest direct
measurements of galaxy masses of e.g. \citet{PerRieVil08}. 
Their published stellar mass densities were multiplied by a factor
$0.61$ since they were estimated with a \citet{Sal55} IMF, which
corresponds to the difference in mass-to-light ratios in the K
band. It is clear that, at all redshifts probed, the range of
SF histories that result from the inversion technique are
consistent with the measured stellar mass density. This consistency
indirectly reinforces the likelihood that the inversion technique spans a reasonable range of possible histories.

Finally, we note that after submission of the present paper, an
erratum was published by Hopkins \& Beacom (2006) in which they
recognize that their computation was erroneous and that, contrary to
their initial claim, the two histories do not exhibit any
inconsistency, apart from possibly at the largest redshifts. Hence their
study is now consistent with our finding, which is not surprising
since our SFR history globally agrees with their compilation (see
Fig.~\ref{figure:sfrdmodel_and_hopkins}).
   
   \begin{figure}[!tbf]
   \includegraphics[width=0.49\textwidth]{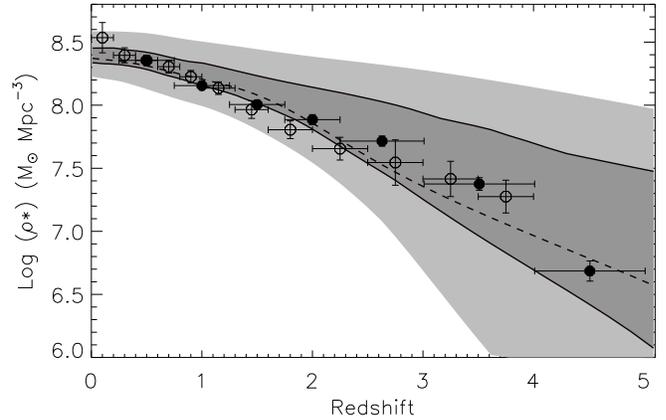}\\
   \caption{Range of allowed evolutions of the stellar mass density
   computed with the range of cosmic SF histories of
   Fig.~\ref{figure:sfrdmodel}. The dashed lines corresponds to the
   best-fit to the IR counts. The data points with error bars are
   measurements from \citet{PerRieVil08} (empty circles) and
   \citet{ElsFeuHop08} (filled dots).  The realistic \citet{BalGla03} IMF
   is used both for the data points and to derive the stellar mass
   density from the SFR density.}
   \label{figure:rhostarmodel} 
   \end{figure}

%


\section{Discussion and conclusions}

\label{section:conclusion}

This paper presents measurements of the evolving infrared luminosity
function and of the corresponding cosmic SFRD
using a non-parametric inversion of the galaxy counts in the mid and far
infrared.

For the first time, we have \emph{exhaustively}  derived the \emph{range} of
possible evolutions for these quantities with a non-parametric inversion
technique. The input data that were used simultaneously to derive these
set of allowed models cover a wide range of wavelengths: deep infrared
counts observed at various wavelengths (from 15 to 850\um{}), the cosmic
infrared background measurements, and optionally the low-redshift
measurements of the IR luminosity function that are derived from the
24\um{} fluxes. We derived from this inversion the allowed range of
SF histories, together with the range of stellar mass
density evolutions that are consistent considering all this
multi-$\lambda$ data. This approach is to be contrasted with previous
modeling works that were based on predictions from a single preferred
model: a \emph{range} of models is given here.

The inversion technique does not use any redshift information as 
input, although such an option can be (and has been) considered
through a prior constructed on the low-redshift LF. Despite this
arguably questionable lack of information about the redshift of the
sources, the inversion technique recovers  the known
redshift distributions surprisingly well up to $z=2$. The reason for this success
probably lies in the very broad basis of wavelengths used in the
inversion: 15\um{}, 160\um{}, and 850\um{} (and intermediate
wavelengths) do not probe the same populations of galaxies at the same
redshifts because of very different k-corrections. The uncertainties
inherent in the library of SEDs that we used seem to be masked, at
first order, by the extent of the data set that we considered.

We find new constraints for the SFRD and its decomposition.  Our method
shows that the IR downsizing must be at work, even though {\sl only} the
IR counts are considered. Quantitatively, we are in good agreement with
direct measurements of the SFRD at low and high redshifts. Again the
clear advantage of our approach is its exhaustivity: the range of
possible SF histories inferred from the inversion does not
suffer from incompleteness, in contrast to surveys based on
spectroscopic and photometric redshifts.  This range matches 
recent measurements of the evolution of the stellar mass density very well, when a
non-evolving IMF is used and stellar remnants are taken into account, in
contrast to previous works.

Then by comparing the results on the SFRD from the ``direct''
(redshift-based) and ``indirect'' (counts inversion) methods, we showed
that the population of HLIRGS tentatively seen at $z>2$ is excluded, at
least with SEDs similar to local ULIRGS. Either the photometric
redshifts of these
peculiar sources are systematically wrong, or their SED is very
different from the local equivalents. Both reasons contribute probably
to the observed discrepancy.

Moreover, a strong contribution from obscured AGNs to the mid-IR counts,
especially at 24\um{} as shown by \citet{DadAleDic07} cannot be
excluded. For this reason, the objects that seem to be HLIRGS at $z=2-3$
may also be actually obscured AGNs. Solid identifications of AGNs from
far-IR diagnostics will be needed for further modeling work.

The set of galaxy SEDs that we used in the inversion (CE01) was
calibrated at $z=0$ for $\lambda < 100$\um{}. Moreover, the SFR
predictions made with these templates are compatible with radio
estimates up to $z=1$ \citep{ElbCesCha02,AppFadMar04}. Although the
multi-$\lambda$ IR counts are reproduced reasonably well with the
inverted luminosity functions, the 160\um{} counts cannot be
satisfactorily reproduced by \emph{any} SF history
satisfying the rest of the counts. At this wavelength, the counts are
systematically under-produced by models. As noted by LDP03, a
population of objects colder than those included in the CE01 library
is needed to produce these counts. We can therefore decide either on
an evolution of the SEDs that were different -- presumably colder for
the same \LIR{} -- at high redshift, or on a poor calibration of the
CE01 SEDs in this range (which is actually expected).  This strongly
confirms the tentative evidence from previous works on modeling
(LDP03) or observational \citet{PapRudLeF07} grounds. The recent works
of \citet{PopChaAle08} on submillimeter galaxies go in the same
direction: these high-redshift IR galaxies are colder than previously
thought. But only Herschel will enable the characterization of the
dust temperature of these objects at these wavelengths. In fact,
Herschel studies of local galaxies will be very
helpful for high-redshift studies of galaxy evolution.

Moreover, the inversion does not completely take  the
special case of AGNs into account. Implicitly, we assumed that their SEDs are close
enough to the templates of star-forming galaxies represented in the CE01
library. Obviously, the AGNs have different SEDs, with a flatter SED in
the mid and far-IR. However, our results are not very sensitive to
this issue. First, the contribution of AGNs (as identified from X-rays
or optical spectra) to the mid-IR counts is small, especially for faint
fluxes. Therefore, a small correction to the bibliographic counts should
be made, but remains difficult at all wavelengths. Moreover, as noted
above, the global trends derived for the LF or the cosmic SFRD compare
well with independent measurements at low redshift, which seems to
indicate that the inversion is relatively insensitive to the precise shape of the SEDs. Of course,
there are limits to this statement, but we believe that the small
fraction of AGNs found in the counts, combined with this statement, should
only weakly affect our findings.

After checking that the predicted redshift distributions of
the sources making the counts at 16, 24, and 70\um{} are compatible
with the real redshift distributions, we then made predictions for the
future surveys to come with Herschel, SPICA, SCUBA2, and Artemis (see
Appendix~\ref{section:predictions} where we
explain the bivariate distributions in $z$ and \LIR{} of the
sources as a function of flux, and we give the fraction of the CIRB
that will be resolved by future confusion limited surveys).

Finally, we stress that, although the \LFIR{} and the SF
history derived from the inversion actually depend (slightly) on the
library of SEDs chosen for the work, the use of a new, updated, SED
library would be straightforward: unlike previous modeling approaches,
it would not imply a whole new work to refine parameters in order to
obtain a good modeling of the counts.

\begin{acknowledgements}
We wish to thank A. Hopkins for making available his compilation of cosmic
SFRs. We also thank Herv\'e Dole for providing
unpublished results from stacking at 70 and 160\um{}. We are grateful
to E. Daddi, R. Chary, H. Dole, G. Lagache, B. Magnelli, K. Glazebrook, and
I. Baldry for fruitful discussions.
D. Le Borgne and D. Elbaz wish to thank the Centre National d'Etudes Spatiales for his
support. D. Elbaz thanks the Spitzer
Science Center at Caltech University for support. 
We would  also like to thank D.~Munro for
freely distributing his Yorick programming language (available at
\texttt{http://yorick.sourceforge.net/}), which was used during the
course of this work. We also thank E. Thiebaut for his optimization
package \texttt{optimpack} \citep{Thi05}. 
\end{acknowledgements}

\bibliographystyle{aa} 
\bibliography{bib_paper2_fixed}
\appendix
\section{The inverse problem}
\label{s:NP}

\subsection{The model}
As argued in the main text, (Sect. 2.1)
the formal equation  relating the number of counts of galaxies  $\mathcal{N}(\lambda_i,S_k)$ 
with the flux $S_k$ (within $\d S$) at wavelength $\lambda_i$ (within $\d\lambda$)
to the  number of counts of galaxies, $N(z,\LIRmath)$, at redshift $z$ (within $\d z$) and IR luminosity $\LIRmath$ 
(within $\d \LIRmath$) is given by 

\begin{equation}
\mathcal{N}(\lambda_i,S_k) = \!\!\int\!\!\!\int\!\delta_{\rm D}\!\left[
S_k-F(\lambda_i,z,\LIRmath{})\right]\!N(z,\LIRmath)\d z\, \d \LIRmath  \,,\EQN{defNN}
\end{equation}
where $\delta_{\rm D}$  is the standard Dirac function,
 $F$ the flux observed in a photometric band centered on
wavelength $\lambda_i$ of a galaxy at redshift $z$ with a \LIR{}
luminosity:
\begin{equation}
F(\lambda_i,z,\LIRmath{})= A / D_L^2(z)  K(\lambda_i,z,\LIRmath{}) \,.
\end{equation}
Here, $D_L(z)$ is the luminosity distance for an object at redshift $z$
with the standard cosmology used in this paper, $A$ is the
solid angle corresponding to one square degree, and $K$ corresponds to 
 the k-corrections:
\begin{equation}
K(\lambda_i,z,\LIRmath{})= 1/R 
\int_{\lambda_i^{\rm min}}^{\lambda_i^{\rm max}}
  \frac{L^{\LIRmath{}} (\lambda/(1+z))}{1+z} T_i(\lambda)
  \mathrm{d}\lambda\,,
\end{equation}
where $T_i(\lambda)$ is the
transmission curve for the filter centered on $\lambda_i$, $R=\int_{\lambda_i^{\rm min}}^{\lambda_i^{\rm max}}
   T_i(\lambda)  \mathrm{d}\lambda$,  and
$L^{\LIRmath{}}(\lambda)$ is the underlying library of SEDs
(CE01) for which the SED  of a galaxy depends only on its
total luminosity \LIR{}. 

As mentioned in the main text, from the point of view of the conditioning of the inverse problem, it is preferable to reformulate  \Eq{defNN}
in terms of $\mathcal{Z}\equiv \log_{10}(1+z)$,
$\mathcal{S}\equiv \log_{10}(S)$ and $m_{\rm IR}\equiv\log_{10} \LIRmath{}$:
\begin{equation}
{\hat \mathcal{N}}(\lambda_i,\mathcal{S}_k) = \!\!\int\!\!\!\int\! H(\mathcal{S}_k,\lambda_i,\mathcal{Z},m_{\rm IR}){\tilde N}(\mathcal{Z},m_{\rm IR})\d{ \mathcal{Z}} \d{m_{\rm IR}} \,,\EQN{defNN2}
\end{equation}
where the kernel of \Eq{defNN2} reads
\[
H(\mathcal{S},\lambda,\mathcal{Z},m_{\rm IR})\equiv 10^{2.5 \mathcal{S}+\mathcal{Z} +m_{\rm IR}}\delta_{\rm D}\!\left[S-F(\lambda,10^\mathcal{Z},10^{m_{\rm IR}})\right] 
\]
with
\begin{equation}
{\tilde N}(\mathcal{Z},m_{\rm IR})\equiv  N(10^\mathcal{Z},10^{m_{\rm IR}})\,,
\end{equation}
\begin{equation}
{\hat \mathcal{N}}(\lambda_i,\mathcal{S}_k)\equiv{\mathcal{N}}(\lambda_i,10^{\mathcal{S}_k}) 10^{2.5 \mathcal{S}_k} \,.
\end{equation}
Here we have introduced the Euclidian-normalized number count, ${\hat \mathcal{N}}$, by multiplying the number count by the expected $S^{2.5}$ power
law.
\subsection{Discretization}

Let us  project ${\tilde N}(\mathcal{Z},m_{\rm IR})$ onto a complete basis of p $\times$ q functions
 \[ \{e_k(\mathcal{Z})   e_l(m_{\rm IR})\}_{  j=1,\ldots,p   \  l=1,\ldots,q}\,,\]   of  finite
(asymptotically  zero) support, which  are chosen here to be piecewise constant step functions:
\begin{equation}
{\tilde N}(\mathcal{Z},m_{\rm IR}) =  \sum_{j=1}^{p} \sum_{l=1}^{q}   n_{jl}  \,  \, e_{j}(\mathcal{Z})
e_{l}(m_{\rm IR}) , \EQN{decomp}
\end{equation}
The parameters  to  fit  are  the  weights $n_{jl}$.
Calling  $\M{X}=\{ n_{jl}\}_{j=1,..p,l=1,..q}$
(the   $ p\times q$   parameters)  and 
$\M{Y}=\{ {\hat \mathcal{N}}(\lambda_i,\mathcal{S}_k)\}_{i=1,..r,k=1,..s}$ 
(the  $r \times s$  measurements), \Eq{defNN2} 
then becomes formally
\begin{equation}
\M{Y}= \M{M}\cdot \M{X} \, , \EQN{yax}
\end{equation}
where $\M{M}$ is a $(r,s)\times (p,q)$ matrix with entries given by
\[
 M_{i k j l}\!\!= \!\! \left\{  {\int \!\!\!  {\int}  \, e_{j}(\mathcal{Z})
e_{l}(m_{\rm IR}) H(\mathcal{S}_k,\lambda_i,\mathcal{Z},m_{\rm IR}) \d{\mathcal{Z}}\d{m_{\rm IR}}}
\right\}_{i k j l}\,  . \EQN{eqnNP}
\]
\subsection{Penalties}
Assuming  that  the noise  in   ${\hat \mathcal{N}}$ can  be
approximated to be Normal, we can estimate the error between the measured
counts and the non-parametric  model by
\begin{equation}
\R{L}_\R{}(\M{X}) \equiv \chi^2(\M{X}) = 
		{({\M{Y}} - \M{M}\mdot \M{X})}^\bot \mdot \M{W}
		\mdot ({\M{Y}} - \M{M}\mdot \M{X}) \,, \EQN{Lquad}
\end{equation}  where 
the weight  matrix $\M{W}$ is  the inverse of  the covariance matrix  of the
data (which is diagonal for  uncorrelated noise with diagonal elements equal
to one over the data variance).
Since we are interested here in a non-parametric inversion,
the  decomposition in  \Eq{decomp} typically  involves many  more parameters
than  constraints,  such that  each  parameter  controls  the shape  of  the
function, ${\tilde N}$,  only   locally. 
As mentioned in the main text, some trade-off must therefore be found between the
level of  smoothness imposed  on the  solution in order  to deal  with  the 
artefacts induced by the ill-conditioning, on the one hand, and the level of fluctuations consistent with the
amount of  information in the  counts, on  the other hand. Between  two solutions yielding equivalent likelihood, the
smoothest  is  chosen on the basis of the quadratic penalty:
\begin{equation}
 	\R{R}_\R{}(\M{X}) = \T{\M{X}} \mdot \M{K} \mdot \M{X}\,, \EQN{Pquad}
\end{equation} where $\M{K}$ is a positive definite matrix, which is chosen 
so that R in \Eq{Pquad} should be non zero when $\M{X}$ is strongly varying as a function
of its indices. In practice, we use a square Laplacian penalization \citet[D2
norm as defined by Eq. 30 of][]{OcvPicLan06}. Indeed, a Tikhonov
penalization does not explicitly enforce smoothness of the solution,
and a square gradient penalization favors flat solutions that are
unphysical in our problem.  

As mentioned in the main text, for a range of redshifts, a direct
measurement, $\mathbf{X}_0$, which can be used as a prior for $\tilde
N$, is available. We may therefore
 add as a supplementary constraint that
\begin{equation}
\R{P}_\R{}(\M{X}) = 
		{({\M{X}} - \M{X}_0)}^\bot \mdot \M{W}_2
		\mdot ({\M{X}} -  \M{X}_0)  \EQN{Lquad2}
\end{equation} 
should remain small,
where 
the weight  matrix, $\M{W}_2$, is  the inverse of  the covariance matrix  of the
prior, $\M{X}_0$, and should be non zero over the appropriate redshift range.
In  short,  the penalized non-parametric  solution 
 of  \Eq{yax}  accounting for both penalties  is found  by
minimizing the so-called objective function
\begin{equation}
Q(\M{X})=L(\M{X})+\lambda\,R(\M{X})+\mu\,P(\M{X}) \,,\EQN{objectif}
\end{equation} 
where $L(\M{X})$, $R(\M{X})$, and $P(\M{X})$ are the  likelihood and  regularization terms given  by \Eqs{Lquad}, 
\Ep{Pquad}, and \Ep{Lquad2},
respectively.
The  Lagrange  multipliers  $\lambda, \mu\geq0$  allow  us  to  tune  the  level  of
smoothness of the solution (in practice, we set $\lambda=0.02$ for the reasons given below) and the requirement that $\M{X}$ should remain close to its prior 
for the range of redshifts for which data is available.  
The introduction  of the  Lagrange multipliers  is
formally justified by our wanting to minimize  the objective function 
$Q(\M{X})$, subject
to   the    constraint   that   $L(\M{X})$ and $P(\M{X})$   should   fall  in   the   range
$N_\R{data}\pm\sqrt{2\,N_\R{data}}$ and
$N_\R{param}\pm\sqrt{2\,N_\R{param}}$, respectively.

The minimum of the objective function, $Q(\M{X})$,  given by \Eq{objectif}
  reads formally as
\[
	\hat \M{X} = (\T{\M{M}} \mdot \M{W} \mdot \M{M}\!+ \lambda \, \M{K}\!+ \mu \, \M{W}_2)^{-1}
	\mdot \left( \T{\M{M}} \mdot \M{W} \mdot \M{Y}\!+\mu \M{W}_2\mdot \M{X}_0\right) \,.
\label{e:Q-quad-solution} 
\]
This equation clearly shows that the solution tends towards $ \M{X}_0$
when $\mu \rightarrow \infty$, while the smoothing Lagrange  multiplier, $\lambda$, damps counterparts of  the components of $\M{Y}$ corresponding to the
higher singular vectors of $\M{M}$ \citep{OcvPicLan06}. 
When  dealing with  noisy datasets,  the non-parametric  inversion technique
may  produce negative  coefficients  for 
the  reconstructed luminosity  function.  To  avoid  such effects,
positivity is imposed  on those coefficients $n_{jl}$ in \Eq{decomp},
see for instance \citet{OcvPicLan06} or \citet{PicSieBie02}. In practice, the minimum of the objective 
function is found iteratively, using \texttt{optimpack} \citep{Thi05}.
The relative weight on the likelihood and the two penalties is chosen
so that the three quantities have a comparable contribution
to the total likelihood after convergence. This corresponds to a
reasonably smooth variation in the LF both in redshift and \LIR{}, and
imposes a solution that is always within 1$\sigma$ of the observed
low-redshift LF, $\M{X}_0$, when $\mu$ is not set to zero in \Eq{objectif}.

\section{Test of robustness}
\label{section:robustness}
To quantify the confidence level of the inversion technique,
we test its robustness.  Starting from an arbitrary LF, we produce IR
counts in the bands and flux ranges corresponding to the observations
from this LF. Then, we add some random Gaussian noise to the simulated
counts, using the real uncertainty on the observations as the $\sigma$
of the error distribution for each flux bin. Finally, we apply the
inversion technique described in Sect. 2.1 to these noisy counts and
obtain an output LF.

\begin{figure}
  \includegraphics[width=0.49\textwidth]{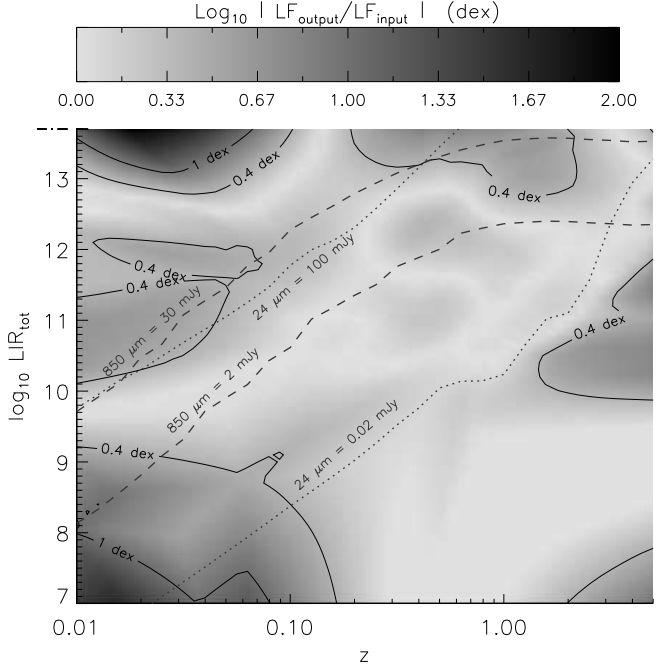}
  \caption{Estimated robustness of the LF inversion used in this paper. The relative
  difference between the input LF and the recovered LF
  (when a realistic noise is added to the corresponding input counts)  is larger for
  darker parts of the diagram. This difference is relatively small ($<$0.4~dex) in the region
  of the z-L space effectively constrained by observations: the dotted
  and dashed lines correspond to the extreme fluxes considered at 24\um{}
  and 850\um{}, respectively, for this study. See main text for details.}
  \label{figure:lfcomp}
\end{figure}

The comparison of the input and output LFs is shown in
Fig.~\ref{figure:lfcomp}. The error on the absolute difference in
$\log_{10} LF_{\rm in} - \log_{10} LF_{\rm out}$ is represented in
gray levels and contours. The difference is generally less than
0.4~dex (factor 2.5) in the range where the LF can be constrained from
the observed counts (range of the $z$-L plane encompassed by the
dashed lines). A noticeable exception is the very low-redshift range
($z<0.1$), which corresponds to bright sources. For such large fluxes,
the considerable noise in the observed counts produces large errors on
the recovered LF. At high redshift, recall that the ultra-luminous
population of galaxies appears as rare and very bright objects, in a
flux range where the number counts are poorly known.

\section{Model predictions for Herschel}
\label{section:predictions}
In Sect.~\ref{section:results}, we have inverted the known IR counts to
obtain constraints on the evolving total IR LF. We have seen that the
LF obtained through this inversion is realistic and matches most of
the recent observations (counts, CIRB, Mid-IR LF at low redshift). 
Then, in Sect.~\ref{section:lfir}, we have shown how we can measure
directly a part of this LF with a good confidence and 
that the LF resulting from the inversion is in good agreement with
this solid measurement, validating the LF obtained by this empirical
modeling approach. In this section, we use the median LF obtained from
the inversions to predict some counts which should be observed with
future observations in the  FIR with Herschel or SCUBA2.

\begin{figure*}
  \includegraphics[width=0.90\textwidth]{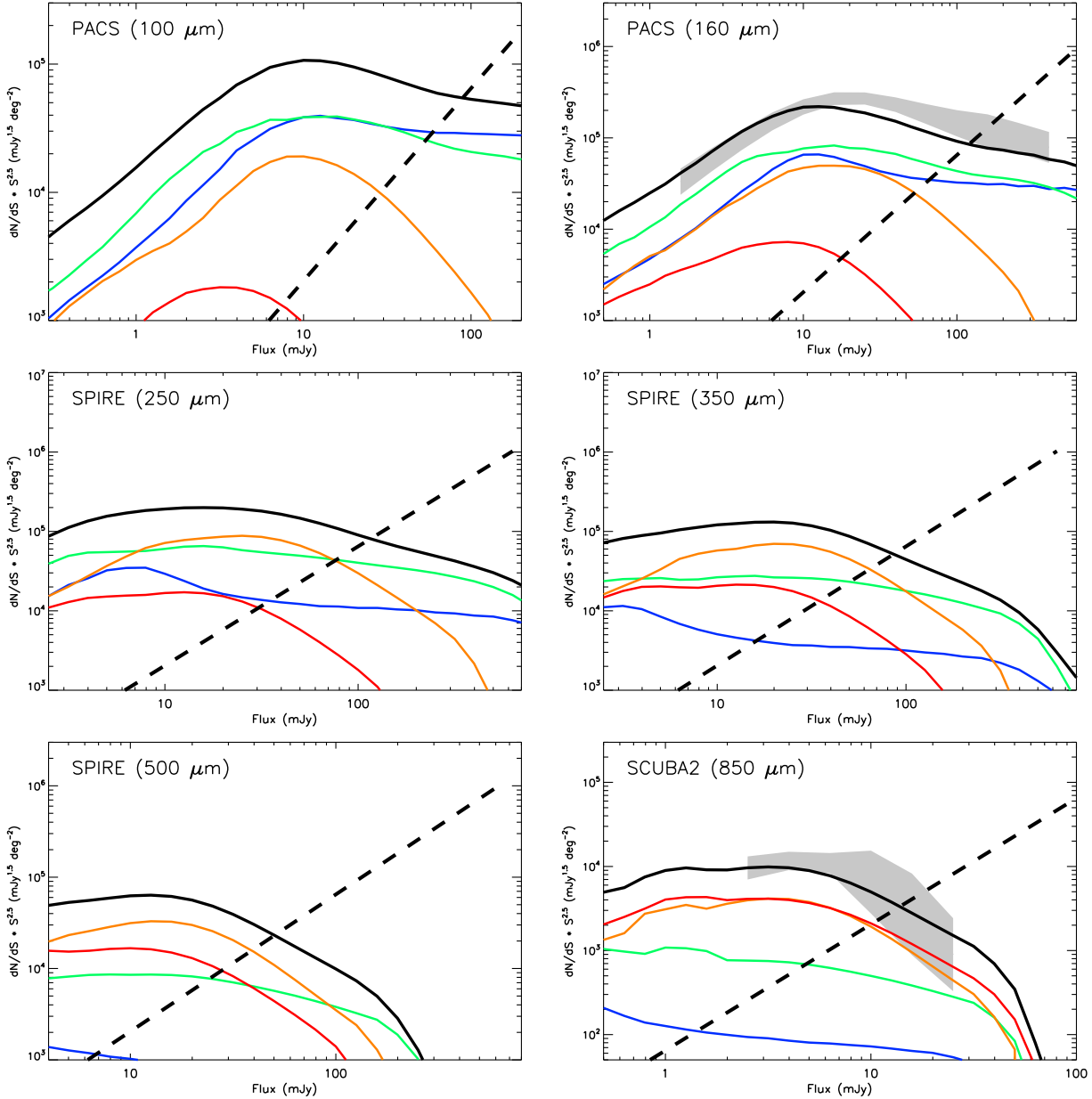}
  \caption{ Counts predicted  from the inversion in the far-infrared and sub-mm (solid line). The counts are
    decomposed in redshift bins (blue=$z<0.5$; green=$0.5<z<1.5$;
    orange=$1.5<z<2.5$; red=$z>2.5$). The oblique dashed line corresponds to
    the limit in statistics due to the smallness of a field like GOODS
    North+South or 0.07 square degrees: less than 2 galaxies per
    flux bin of width $\delta \log$F=0.1~dex are expected below
    this limit.}
  \label{figure:zdists_herschel}
\end{figure*}

\begin{figure*} 
    \includegraphics[width=0.99\textwidth]{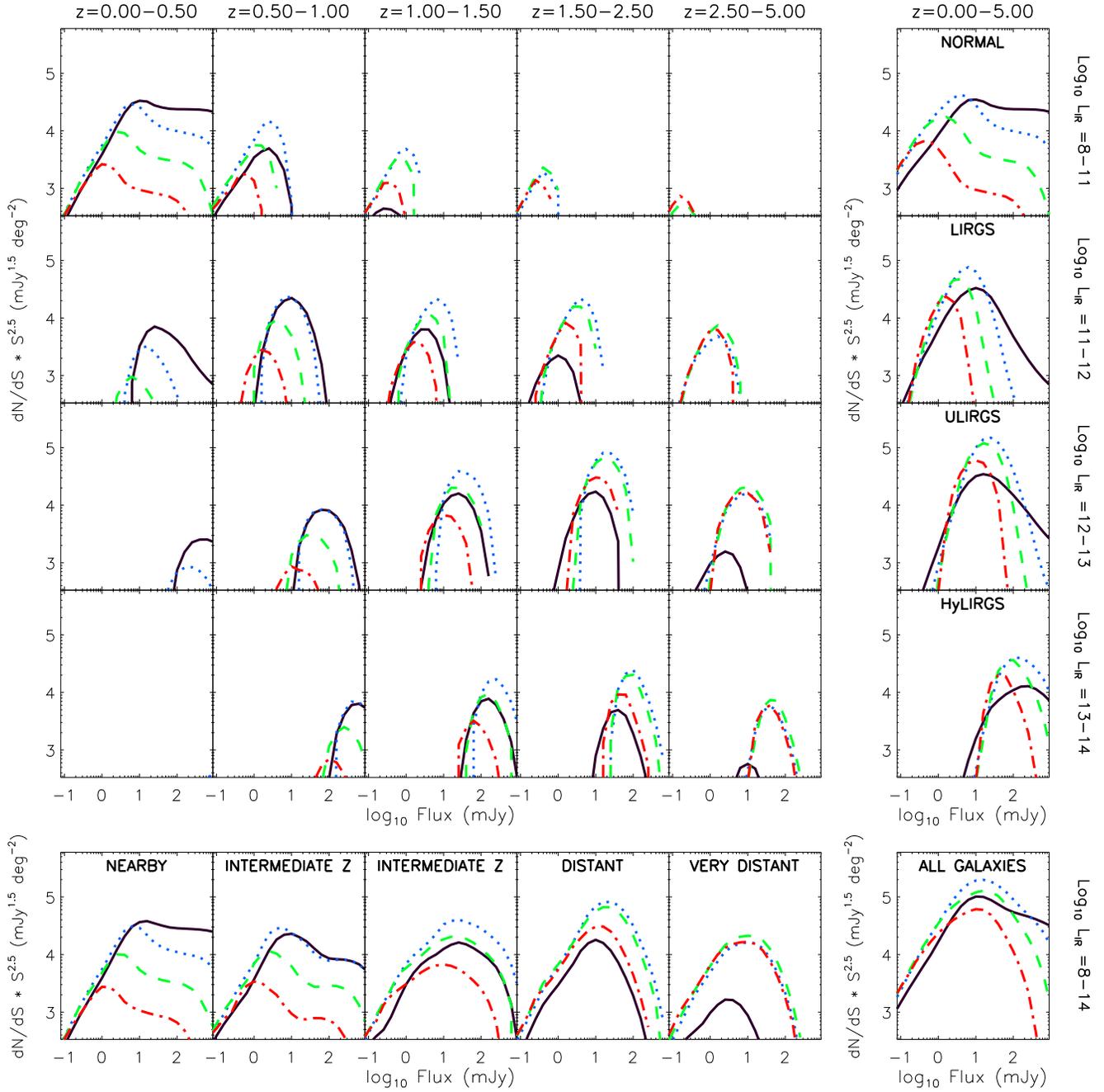}
    \caption{Differential counts predicted from the non-parametric
    inversion for future Herschel observations: PACS 100\um{} (solid
    black), SPIRE 250\um{} (dotted blue),
    SPIRE 350\um{} (dashed green), SPIRE 500\um{} (dot-dashed red) decomposed simultaneously in
    redshift and \LIR{}.}
    \label{figure:H2}
  \end{figure*}

At the time of publication, several new facilities
are in preparation to observe the Universe in the far-IR to sub-mm
regimes. The differential counts (normalized to Euclidean) at
wavelengths ranging from 16 to 850\,$\mu$m, which we derived from the
inversion technique, are presented in Fig.~\ref{figure:zdists_herschel}. The
separation of the contribution of local, intermediate, and distant
galaxies in different colors illustrates the expected trend that
larger wavelengths are sensitive to higher redshifts, hence the relative
complementarity of all IR wavelengths. There will be a bias towards
more luminous and distant objects with increasing wavelength,
illustrated here for the Herschel passbands (see
Fig.~\ref{figure:H2}), but this may be used to pre-select the most
distant candidates expected to be detected only at the largest
wavelengths. In the following, we discuss the predictions of the
inversion technique for those instruments, as well as their respective
confusion limits, which is the main limitation of far-IR extragalactic
surveys.

\begin{figure} 
    \includegraphics[width=0.49\textwidth]{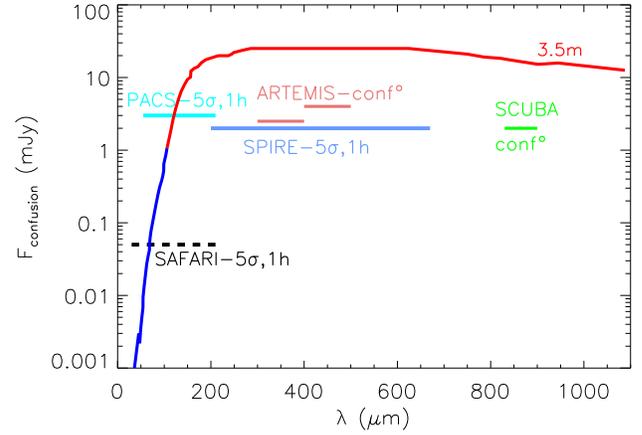}
    \caption{Confusion limit for a 3.5 meter telescope. The
    5$\sigma$-1hour limits of SPICA-SAFARI (50\,$\mu$Jy,
    33-210\,$\mu$m, dashed line), Herschel PACS (3mJy, 55-210\,$\mu$m,
    light blue line) and SPIRE (2 mJy, 200-670\,$\mu$m, blue line) are
    shown together with their wavelength ranges. The blue part of the
    curve is determined by the source density criterion (i.e. the
    requirement to have less than 30\% of the sources closer than
    0.8$\times$FWHM), the red part is defined by the photometric
    criterion, i.e. sources must be brighter than 5 times the rms due
    to very faint sources below the detection limit.}
    \label{figure:conf_limit}
  \end{figure}

The ESA satellite Herschel is scheduled to be launched within the next
year, while the next-generation IR astronomical satellite of the
Japanese space agency, SPICA, is scheduled for 2010, with a
contribution by ESA under discussion, including a mid-IR imager named
SAFARI.  Both telescopes share the same diameter of 3.5 meters, but
the lower telescope temperature of SPICA, combined with projected
competitive sensitivities, will make it possible to reach confusion
around 70\,$\mu$m (where Herschel is limited by integration time). The
5$\sigma$-1hour limits of the instruments SAFARI onboard SPICA
(50\,$\mu$Jy, 33-210\,$\mu$m, dashed line), PACS (3mJy,
55-210\,$\mu$m, light blue line) and SPIRE (2 mJy, 200-670\,$\mu$m,
blue line) onboard Herschel are compared in Fig.~\ref{figure:conf_limit}
to the confusion limits that we derive from the best-fit model of the
inversion, at all wavelengths between 30 and 850\,$\mu$m, assuming the
the confusion limit definition given below.

  \begin{figure} 
    \includegraphics[width=0.49\textwidth]{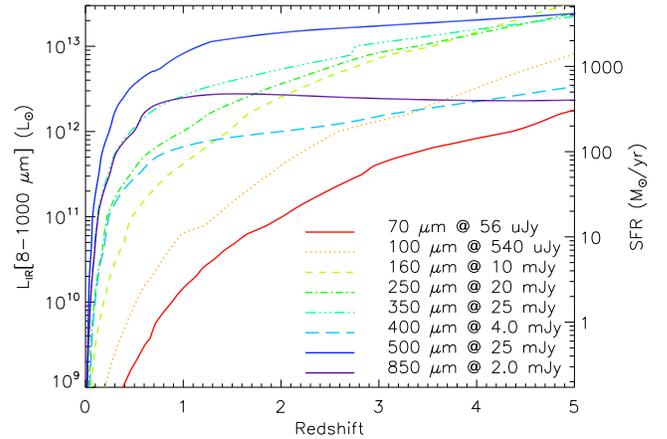}
    \caption{Detection limits for confusion limited surveys from 70 to
    850\,$\mu$m. The curves show the minimum IR luminosity
    (8-1000\,$\mu$m), or equivalently SFR (=L$_{\rm
    IR}\times1.72~10^{-10}$), that can be detected for a star-forming
    galaxy assuming that it has an SED similar to the Chary \& Elbaz
    (2001) ones. The 70, 100, 160, 250, 350 and 500\,$\mu$m limits
    correspond to a 3.5 m telescope diameter, such as Herschel or
    SPICA, while the 400\,$\mu$m is for a 12 m class telescope such as
    APEX (e.g. ARTEMIS, we show the average between the two bands at
    350 or 450\,$\mu$m to avoid confusion with Herschel) and the
    850\,$\mu$m is for a 15 m telescope as the JCMT (SCUBA).}
    \label{figure:Lirz}
  \end{figure}

The definition of the confusion limit is not trivial, in particular
because it depends on the level of clustering of galaxies; the
optimum way to define it would be to perform simulations to compute
the photometric error as a function of flux density, and then decide
that the confusion limit is e.g. the depth above which 68\,\% of the
detected sources are measured with a photometric accuracy better than
20\,\%. In the following, we only consider a simpler approach that
involves computing the two sources of confusion that were discussed
in \citet{DolLagPug03}: \\
- the photometric confusion noise: the noise produced by sources
fainter than the detection threshold. The photometric criterion corresponds 
to the requirement that sources are detected with an
S/N(photometric)$>$5.\\
- the fraction of blended sources: a requirement for the quality of
  the catalog of sources will be that less than $N$\,\% of the sources
  are closer than 0.8$\times$FWHM, i.e. close enough to not be
  separated. \\

We tested various levels for $N$ and found that $N=30$\,\% was
equivalent to the above requirement that 68\,\% of the detected
sources are measured with a photometric accuracy better than 20\,\%
using realistic simulations in the far IR for Herschel. We therefore use the value  $N=$30\,\%. The confusion limit is then
{\sl defined} as the flux density above which both criteria are
respected. As a result, it is found that the main limitation is the
fraction of blended sources at $\lambda$=50-105\,$\mu$m (blue part of
the curve in Fig.~\ref{figure:conf_limit}) and the photometric confusion
noise below and above this range, i.e. at $\lambda$= 33-50 and
$\lambda$=105-210\,$\mu$m (red parts of curve in
Fig.~\ref{figure:conf_limit}). As a result of their smaller beam, shorter
IR wavelength are more efficient at detecting faint star-forming
galaxies than longer ones (see Fig.~\ref{figure:Lirz}). This is at the
expense of observing farther away from the peak of the far IR
emission, which implies larger uncertainties on the derivation of the
total IR luminosity due to the uncertain dust temperature.

We note that the confusion limit for a 3.5m-class telescope, such as
Herschel, is ten times more than the depth it can reach in one hour
(5$\sigma$). With a source density of 12.8 sources per square degree
at the 500\,$\mu$m confusion limit (25 mJy), or equivalently one
source in a field of 17 arcmin on a side, this shows that the best
strategy at this wavelength is to go for very large and moderately
shallow surveys, in order to identify the rare and very luminous distant
galaxies.

\begin{table*}

\begin{center}
\caption{Fraction of the CIRB resolved by confusion-limited Herschel
  surveys}
\label{table:cirb}
\begin{tabular}{lcccccc}
\hline
\hline
 & PACS 70 \um{} & PACS 100 \um{} &  PACS 160 \um{} &  SPIRE 250 \um{}
&  SPIRE 350\um{} &  SPIRE 500\um{} \\
\hline
F$_\mathrm{confusion}$ (mJy) & 0.056& 0.54& 10.0& 20.0& 25.1& 25.1 \\
IGL$^\mathrm{a}$ @ F$_\mathrm{confusion}$ & 9.14& 12.3& 6.27& 1.98& 0.55& 0.058 \\
CIRB$^\mathrm{a}$ & 9.51& 14.0& 15.3& 10.3& 5.47& 2.29 \\
\% CIRB resolved& 96.1 \%& 87.8 \%& 41.1 \%& 19.1 \%& 10.0 \%& 2.5 \% \\
\hline

\end{tabular}
\begin{list}{}{}
\item[$^{\mathrm{a}}$] Units: nW.m$^{-2}$.sr$^{-1}$
\end{list}
\end{center}
\end{table*}

For comparison, we also illustrated the ground-based capacity of
ARTEMIS built by CEA-Saclay which will operate at the ESO 12
m-telescope facility APEX (Atacama Pathfinder EXperiment) at 200, 350
and 450\,$\mu$m and SCUBA-2 that will operate at the 15 m telescope
JCMT at 450 and 850\,$\mu$m. To avoid confusion between all
instruments, we only show the average wavelength 400\,$\mu$m for a 12
m-class telescope and 850\,$\mu$m for a 15 m-class telescope
(Fig.~\ref{figure:conf_limit}). Although the confusion limit in the
850\,$\mu$m passband is ten times below that of Herschel at the
longest wavelengths, this band is not competitive with the
$\sim$400\,$\mu$m one, which should be priorities for ARTEMIS and
SCUBA-2 for the study of distant galaxies, or with the 70 and
100\,$\mu$m ones for a 3.5 m space experiment such as SPICA and
Herschel, for redshifts below $z\sim$5. We also note that only in
these two passbands will the cosmic IR background be resolved with
these future experiments (see Table~\ref{table:cirb}), which suggests
that a larger telescope size should be considered for a future
experiment to observe the far IR Universe above 100\,$\mu$m and below
the wavelength domain of ALMA. We did not mention here ALMA since it
will not be affected by these confusion issue:  due to its very
good spatial resolution, it will be limited to either small ultradeep
survey, hence missing rare objects or follow-ups of fields observed
with single dish instruments, e.g. ARTEMIS. Finally, the JWST that
will operate in the mid IR will be a very powerful instrument for probing
the faintest star-forming galaxies in the distant Universe, but
predictions are difficult to produce at the present stage since it has
already been found that extrapolations from the mid to far IR become less
robust already at $z\sim$2 \citep[e.g.][]{DadDicMor07,PapRudLeF07,PopChaAle08}.

\end{document}